\documentclass[aps,prl,reprint,superscriptaddress,amsmath,amssymb]{revtex4-2}

\usepackage{graphicx}
\usepackage{bm}
\usepackage{booktabs}
\usepackage{hyperref}

\newcommand{\dd}{\mathrm{d}}
\newcommand{\eq}{\mathrm{eq}}
\newcommand{\pvec}{\bm p}
\newcommand{\xperp}{\bm x}
\newcommand{\Etot}{E_{\mathrm{tot}}}
\newcommand{\eff}{\mathrm{eff}}

\begin{document}

\title{Beyond Viscosity Matching: Microscopic Relaxation in Anisotropic Flow}

\author{Olga Soloveva}
\email{soloveva@itp.uni-frankfurt.de}
\affiliation{GSI Helmholtzzentrum f\"ur Schwerionenforschung GmbH, Planckstra\ss e 1, 64291 Darmstadt, Germany}
\affiliation{Institut f\"ur Theoretische Physik, Johann Wolfgang Goethe-Universit\"at, Max-von-Laue-Str.~1, 60438 Frankfurt am Main, Germany}
\affiliation{Helmholtz Research Academy Hessen for FAIR (HFHF), GSI Helmholtz Center for Heavy Ion Physics, Campus Frankfurt, 60438 Frankfurt am Main, Germany}

\begin{abstract}
When a gas of scattering particles is disturbed, each angular harmonic of its
velocity distribution decays at its own rate. Hydrodynamics, and with it the
shear viscosity, depends on only one of these rates, the one for the quadrupole harmonic. A system that scatters only a few times before it stops
interacting remembers all of them. This is the situation of small collision systems such as $p+p$, $p+$Pb and O$+$O, where kinetic models tuned to the same viscosity nevertheless predict different anisotropic flow. We show that this
difference is not a model uncertainty but a measurement of the remaining rates: to first order in the number of collisions, ratios of flow responses equal ratios of relaxation rates, independently of the initial geometry. For the QCD
collision kernel, where the rates also depend on momentum, we obtain the full spectrum of relaxation modes in closed form and show that dilute flow, viscosity, and late-time decay probe three different averages of it. A momentum-resolved kinetic solver confirms that the angular hierarchy is lost first at soft momenta and that the flow response settles at a value no single-rate model produces. The same structure governs two-dimensional electron and atomic Fermi gases.
\end{abstract}

\maketitle

\textit{Introduction.}
Collective flow is observed in collision systems as small as $p+p$, $p+$Pb,
O$+$O and Ne$+$Ne \cite{Nagle:2018nvi,Schenke:2021mxx,Brewer:2021kiv,Giacalone:2024luz,Krupczak:2026oo}.
Such systems live at intermediate \emph{opacity}: a typical particle scatters
only a few times before the fireball has expanded away, so the Knudsen number is
of order one. Relativistic kinetic theory is the natural language for this
regime \cite{He:2015hfa,Bhatnagar:1954zz,Anderson:1974nyl,Arnold:2002zm}, and
it is usually solved in one of two ways: with the relaxation-time
approximation (RTA), where every deviation from equilibrium decays at one rate
$1/\tau_R$, or with QCD effective kinetic theory (EKT), where the full
small-angle scattering kernel is kept. The two approaches are calibrated to the
same shear viscosity $\eta/s$, and yet they predict different flow at
intermediate opacity. This difference is currently treated as a model
uncertainty \cite{Kurkela:2018qeb,Kurkela:2019kip,Ambrus:2021fej,Ambrus:2022qya,Kurkela:2018vqr,Kurkela:2018wud,Bernhard:2019bmu}.

The purpose of this Letter is to show that the difference is not an
uncertainty but a measurement. The argument has two steps. First, shear
viscosity is a property of a single eigenvalue of the collision operator, the
one governing the $l=2$ angular harmonic, whereas the flow of a dilute system
is generated by \emph{all} angular eigenvalues, and we prove that it measures
them in a geometry-independent way. Second, the actual QCD kernel does not
have a single rate per harmonic: soft and hard partons carrying the same
angular deformation relax at different rates. We compute the resulting
relaxation spectrum exactly and show that flow, viscosity, and late-time decay
weight it differently. Matching $\eta/s$ alone therefore leaves the flow of
small systems undetermined by construction, and momentum-differential flow
fixes what is missing.
This changes how kinetic models should be compared and how the theoretical
uncertainty of small-system flow should be quoted: not as a band between two
calculations tuned to the same $\eta/s$, but as the range of a few additional
numbers, the low-lying moments of the relaxation spectrum, that flow data can
constrain.

Throughout, ``mode'' means an eigenmode of the collision operator, not a mode
of the initial-state geometry \cite{Borghini:2022vna,Krupczak:2025mbm}; the
closest analogue of the mechanism is the tomographic transport of
two-dimensional electron fluids \cite{Ledwith:2019a,Lucas:2018scl,Gurzhi:1968ur},
to which we return at the end.

\textit{Setting.}
We consider massless partons in the transverse plane, the two-dimensional model
widely used to study the onset of flow \cite{Borghini:2018xum,Roch:2020zdl,Bachmann:2022cls}.
Since the collision kernels of interest act on the momentum angle $\phi$, we
integrate out the modulus with energy weight, $\Phi(\xperp,\phi,\tau)=\int
\dd p\,p^2 f$, and expand in circular harmonics, $\Phi=\sum_l\Phi_l(\xperp,\tau)e^{il\phi}$.
The Boltzmann equation is $(\partial_\tau+\bm v\cdot\nabla_\perp)\Phi=C[\Phi]$,
and the energy-weighted flow coefficient is $V_n\propto\int\dd^2x\,\Phi_{-n}$.
We define the response $\kappa_n\equiv v_n/\varepsilon_n$, the ratio of the
final flow harmonic to the initial eccentricity of the same order, and we
quantify opacity by a dimensionless number $g$, normalized so that the $l=2$
harmonic of a thermal distribution relaxes at rate $g$ per unit time in units of
the system size; $g$ is an inverse Knudsen number.

Any collision operator that is rotation invariant in the local rest frame is,
once linearized around the Landau-matched equilibrium $\Phi^{\eq}$, diagonal in
these harmonics. The most general such \emph{scalar} model is
\begin{equation}
 C[\Phi]= -\sum_l\gamma_l(\xperp,\tau)
 \bigl(\Phi_l-\Phi_l^{\eq}\bigr)e^{il\phi},
 \qquad \gamma_0=\gamma_{\pm1}=0 .
 \label{eq:scalarC}
\end{equation}
The two vanishing rates are not a choice: energy conservation requires
$\int\dd\phi\,C=0$ and momentum conservation $\int\dd\phi\,\bm v\,C=0$, which
force $\gamma_0=\gamma_{\pm1}=0$. (A naive angular-diffusion operator
$D\partial_\phi^2$ has $\gamma_1=D\neq0$ and violates momentum conservation.)
The standard RTA is the special case $\gamma_l=1/\tau_R$ for all $l\ge2$;
momentum-conserving angular diffusion is $\gamma_l=D(l^2-1)$; a kernel with both
a large-angle and a small-angle component is $\gamma_l=a+bl^2$, and its
hierarchy $\gamma_3/\gamma_2=(a+9b)/(a+4b)$ interpolates between $1$ (RTA) and
$9/4$ (pure diffusion). We call Eq.~\eqref{eq:scalarC} the mode-resolved RTA
(MRTA). Our single structural assumption for the scalar case is that all
harmonics share one spatial profile,
\begin{equation}
 \gamma_l(\xperp,\tau)=\widehat g_l\,W(\xperp,\tau),
 \label{eq:separable}
\end{equation}
with $W$ a common medium field, e.g.\ $W\propto e^\alpha$ with $\alpha=1$ for a
rate proportional to the density and $\alpha=1/3$ for a conformal rate
proportional to $T$.

\textit{What flow measures for a scalar spectrum.}
In a dilute system most particles scatter at most once, so it is natural to
organize the solution by the number of collisions, $\Phi=\Phi^{(0)}+\Phi^{(1)}+\dots$. The
free-streaming background $\Phi^{(0)}(\xperp,\phi,\tau)=\Phi_0(\xperp-\bm v t,\phi)$
generates no flow at all: its spatial integral is a constant of the motion,
hence $V_n^{(0)}=0$ for every $n\ge1$ at every time. All flow is collisional.
Inserting the first correction into $V_n$, shifting the integration variable
along the free-streaming characteristic (unit Jacobian), and using the
diagonality of $C$, one finds

\begin{equation}
 V_n^{(1)}=-\widehat g_n\,\mathcal G_n[E_0,W],
 \label{eq:onehit}
\end{equation}
with the geometry functional
\begin{equation}
  \mathcal G_n[E_0] \;\equiv\; \frac{2\pi}{\Etot}
  \int \dd\tau\, \dd^2x\;\;
  W\,\bigl[\Phi^{(0)}_{-n} - \Phi^{\eq[0]}_{-n}\bigr],
  \label{eq:Gn}
\end{equation}
where $\Phi^{\eq[0]}$
is the equilibrium Landau-matched to the free-streaming
background. Everything about the geometry, the initial profile $E_0$, the
choice of $W$, and the equilibrium subtraction, sits in $\mathcal G_n$; the
only place the spectrum enters is the prefactor $\widehat g_n$. Consequently,
for any two spectra $A$ and $B$,
\begin{equation}
 \frac{(\kappa_n/\kappa_m)_A}{(\kappa_n/\kappa_m)_B}
 =\frac{(\widehat g_n/\widehat g_m)_A}{(\widehat g_n/\widehat g_m)_B}.
 \label{eq:ratio}
\end{equation}

In words: at leading order in opacity, the ratio of flow responses of two
harmonics, normalized to the same ratio in a reference model, is exactly the
ratio of the microscopic relaxation rates, whatever the initial geometry. We
denote this double ratio, taken against the flat RTA, by $D_{nm}$; for pure
angular diffusion $D_{32}=9/4$ and for its momentum-conserving version
$D_{32}=8/3$.

Two remarks make the result useful. First, the equilibrium subtraction in
Eq.~\eqref{eq:onehit} is not small (free streaming builds up large local
anisotropies), but it carries the same $\widehat g_n$ as the direct term, so
Eq.~\eqref{eq:ratio} is blind to it. Second, for $\alpha=1$ the direct term of
every odd harmonic vanishes at linear order in the eccentricity, an even--odd
selection rule that reduces, for $n=1$, to the statement that a rigid
translation produces no flow; odd flow at $\alpha=1$ is generated entirely by
the subtraction, again with prefactor $\widehat g_n$.

\textit{Why viscosity does not see this.}
The drift term $\bm v\cdot\nabla_\perp$ couples harmonic $m$ only to $m\pm1$,
so a Chapman--Enskog expansion of the Boltzmann equation to first order in
gradients closes on the $l=\pm2$ channel:
$\gamma_2\,\delta\Phi_{\pm2}=-\tfrac12(\partial_x\mp i\partial_y)\Phi_{\pm1}+O(\nabla^2)$.
The shear stress is built from $\delta\Phi_{\pm2}$, and therefore
\begin{equation}
 \eta\propto\frac{e+P}{\gamma_2},\qquad \tau_\pi=\frac1{\gamma_2}.
 \label{eq:eta2}
\end{equation}
The rates $\gamma_{l\ge3}$ first appear at third order in gradients. Any two
kernels that share $\gamma_2$ therefore share $\eta$ and $\tau_\pi$, and differ
only where the gradient expansion has not yet converged. That regime is the
intermediate opacity of small systems, and Eq.~\eqref{eq:ratio} says what is
measured there.

\textit{From dilute to hydrodynamic.}
Beyond one collision the same $\gamma_n$ that generates $v_n$ also damps it:
along a characteristic, $\Phi_n(\infty)=\int\dd\tau\,S_n(\tau)\,\exp[-\int_\tau^\infty\gamma_n]$
with $S_n$ the gradient source, so $\kappa_n$ is enhanced by $\gamma_n$ at low
opacity and suppressed by $e^{-\int\gamma_n}$ at high opacity, the kinetic
counterpart of the Gurzhi crossover. A deterministic solver with exact
spectral advection and exact per-harmonic relaxation (energy and momentum
conserved to $10^{-14}$; dilute anchors of Eq.~\eqref{eq:ratio} reproduced to
four digits) shows that the signal decays slowly: $D_{32}$ drops from $9/4$ to
$1.92$ at $g=8$, and $|D_{32}-1|<0.1$ is reached only at $g\gtrsim130$.
Sensitivity to the higher modes thus persists over three decades of opacity,
from peripheral $p+p$ to central Pb$+$Pb. In boost-invariant 3D kinematics
the longitudinal expansion mixes the angular label $l$ with the azimuthal
label $n$ and washes the signal out as
$\gamma_n^\eff/\gamma_2^\eff\simeq1+(n^2-4)\Delta^2$ for a small longitudinal
width $\Delta\sim\tau_0/R$ (Supplemental Material): small systems, with
$\tau_0/R=O(1)$, retain the full spectral sensitivity, while large systems
lose it twice, hydrodynamically and through the longitudinal squeeze.

\textit{The QCD kernel is not scalar.}
Equation~\eqref{eq:separable} assigns one number to each harmonic. The
leading-logarithmic limit of the linearized QCD collision operator
\cite{Arnold:2000dr,Arnold:2002zm,Hong:2010at} does not have this form. It is a
Fokker--Planck operator in momentum,
\begin{equation}
 C_{\rm FP}[f]=\varkappa\,\bm\nabla_p\!\cdot
 \left(T\bm\nabla_p f+\widehat{\pvec}f\right),
 \label{eq:FP}
\end{equation}
with $\varkappa$ the momentum-diffusion coefficient ($\sim\hat q$) and $T$ the
bath temperature; its stationary point is the thermal exponential $e^{-p/T}$.
Rotation invariance still makes it diagonal in circular harmonics, but each
harmonic block is now a differential operator in the modulus. Writing
$C_l=-\Lambda_l$,
\begin{equation}
 \Lambda_l=-\varkappa\left[
 \partial_p^2+\Bigl(\frac1p+\frac1T\Bigr)\partial_p
 +\frac1{pT}-\frac{l^2}{p^2}\right].
 \label{eq:block}
\end{equation}
The last term is the angular relaxation rate, and it is momentum dependent,
\begin{equation}
 \gamma_l(p)=\frac{\varkappa\,l^2}{p^2}.
 \label{eq:localrate}
\end{equation}
Soft partons isotropize almost instantly; hard partons barely at all. The
angular index and the momentum modulus are entangled, and no choice of
$\widehat g_lW$ in Eq.~\eqref{eq:separable} reproduces this operator. Our
transport model is Eq.~\eqref{eq:scalarC} with the scalar rates replaced by
the operators $\Lambda_l$ for $|l|\ge2$, the $l=0,\pm1$ channels frozen so that
conservation remains exact (a full conservation projection is compared in the
Supplemental Material and changes nothing below), and the Landau-matched
equilibrium lifted to momentum space, $F^{\eq}\propto p^2e^{-p\,\sigma(\phi)/T}$
with $\sigma=u^\tau-\bm u_\perp\cdot\widehat{\bm v}$. We normalize
$\varkappa=T^2/2$ so that the thermal one-hit $l=2$ rate equals $g$; double
ratios against the flat RTA are then taken at equal opacity.

\textit{The relaxation spectrum in the closed form.}
$\Lambda_l$ is self-adjoint with respect to the weight $p\,e^{p/T}$, and the
Liouville substitution $\psi=(p\,e^{p/T})^{-1/2}\chi$ maps it onto a radial
Coulomb Hamiltonian,
\begin{equation}
 \Lambda_l\simeq\varkappa\left[-\frac{\dd^2}{\dd p^2}
 +\frac{l^2-\tfrac14}{p^2}-\frac{1}{2Tp}+\frac{1}{4T^2}\right],
 \label{eq:coulomb}
\end{equation}
with angular momentum $l-\tfrac12$, charge $1/2T$, and an energy offset
$\varkappa/4T^2$ that is the relaxation floor set by drag alone. This
Fokker--Planck--Coulomb correspondence was recently used by Gavassino to obtain
the quasinormal modes of relativistic Fokker--Planck kinetic theory
\cite{Gavassino:2026zsz}. We apply it to a different sector: the
momentum-conservation-constrained anisotropy blocks $l\ge2$ that carry flow.
The spectrum is hydrogenic, a tower of bound states accumulating at a continuum
edge,
\begin{equation}
 \gamma_{l,k}=\frac{\varkappa}{4T^2}
 \left[1-\frac{1}{(2l+2k+1)^2}\right],\quad k=0,1,\dots,
 \qquad
 \label{eq:spectrum}
\end{equation}
$ \gamma_{\rm cont}\ge\frac{\varkappa}{4T^2},$
with eigenfunctions $\psi_{l,k}\propto p^{\,l}e^{-p/2T}e^{-p/4T\nu}L^{(2l)}_k(p/2T\nu)$,
$\nu=l+k+\tfrac12$. A fine-grid diagonalization of Eq.~\eqref{eq:block}
reproduces Eq.~\eqref{eq:spectrum} to $3\times10^{-4}$, including the Coulomb
degeneracy $\gamma_{l,k}=\gamma_{l+1,k-1}$.

The shape of this spectrum is the central physical point. The $l^2$ hierarchy
of Eq.~\eqref{eq:localrate} lives entirely in the continuum, i.e.\ at soft
momenta. The bound states describe deviations that have escaped to hard
momenta, where relaxation is drag dominated and almost blind to $l$: the whole
flow-carrying tower is squeezed between $0.96$ and $1$ times the edge, and in
particular
\begin{equation}
 \frac{\gamma_{3,0}}{\gamma_{2,0}}=\frac{1-1/49}{1-1/25}=\frac{50}{49}\simeq1.0204,
 \label{eq:late}
\end{equation}
to be compared with $9/4$ for the naive angular-diffusion reading of the same
kernel. Whatever survives many collisions relaxes at an $l$-independent rate.
The mode-resolved hierarchy of the QCD kernel is therefore a transient,
soft-momentum phenomenon.

\textit{Three observables, three averages.}
Equation~\eqref{eq:ratio} has an exact operator analogue. Let the channel-$n$
deviation have momentum profile $h(p)$ (in $f$-space). Because the
$l$ dependence of Eq.~\eqref{eq:block} is exactly $l^2\times\varkappa/p^2$, the
effective rate seen by the energy-weighted response is computable for any $h$:
\begin{equation}
 \gamma_n^{\eff}[h]
 \equiv\frac{\langle p^2|\Lambda_n|h\rangle}{\langle p^2|h\rangle}
 = A[h]+n^2B[h],
 \label{eq:effective}
\end{equation}
with $A=\varkappa(I_1/T-I_0)/I_2$, $B=\varkappa I_0/I_2$, and
$I_m=\int_0^\infty\dd p\,p^mh$. Equivalently, expanding $h$ in the eigenbasis
of Eq.~\eqref{eq:coulomb}, $\gamma_n^\eff=\sum_kw_{nk}\gamma_{n,k}+(\text{continuum})$
with weights given by overlaps of the source with the eigenfunctions. At
leading opacity the QCD kernel is thus indistinguishable from a mixed MRTA
spectrum $a+bl^2$, but with $a$ and $b$ fixed by the hardness of the source.
For a thermal source, $h=e^{-p/T}$, one finds $A=0$ and $\gamma_n^\eff=\varkappa n^2/2T^2$,
so the anchors $9/4$ and $4$ of angular diffusion are recovered exactly. The
Landau-matched subtraction, however, has channel shapes
\begin{equation}
 h_n^{\eq}(p)\propto p^2e^{-pu^\tau/T}I_n(pu_\perp/T),
 \label{eq:landau}
\end{equation}
harder than thermal and explicitly $n$ dependent ($I_n$ is the modified Bessel
function). Its effective eigenvalues are $0.279$ and $0.346$ for $n=2,3$, a
double ratio of only $1.24$, against $1$ and $9/4$ for the direct term. This,
and only this, is where scalar factorization breaks: the two contributions to
$v_n$ no longer carry a common prefactor.

The same block $\Lambda_2$ presents different numbers to different
measurements. In units where the thermal one-hit rate is $1$: the dilute
response measures the direct matrix element, $1$; the exact Chapman--Enskog
solution $p^2e^{-p/T}$ gives
\begin{equation}
 \frac{\eta_{\rm FP}}{\eta_{\rm flat}}=4,
 \label{eq:eta}
\end{equation}
so viscosity measures an inverse spectral moment, $1/4$; and the late-time
decay measures the ground state, $\gamma_{2,0}=0.12$. A flat RTA matched to
$\eta$ therefore has four times the dilute response of the QCD kernel and eight
times its asymptotic decay rate. This is the microscopic reason why viscosity
matching cannot fix intermediate-opacity flow.

\begin{figure*}[t]
 \includegraphics[width=0.96\textwidth]{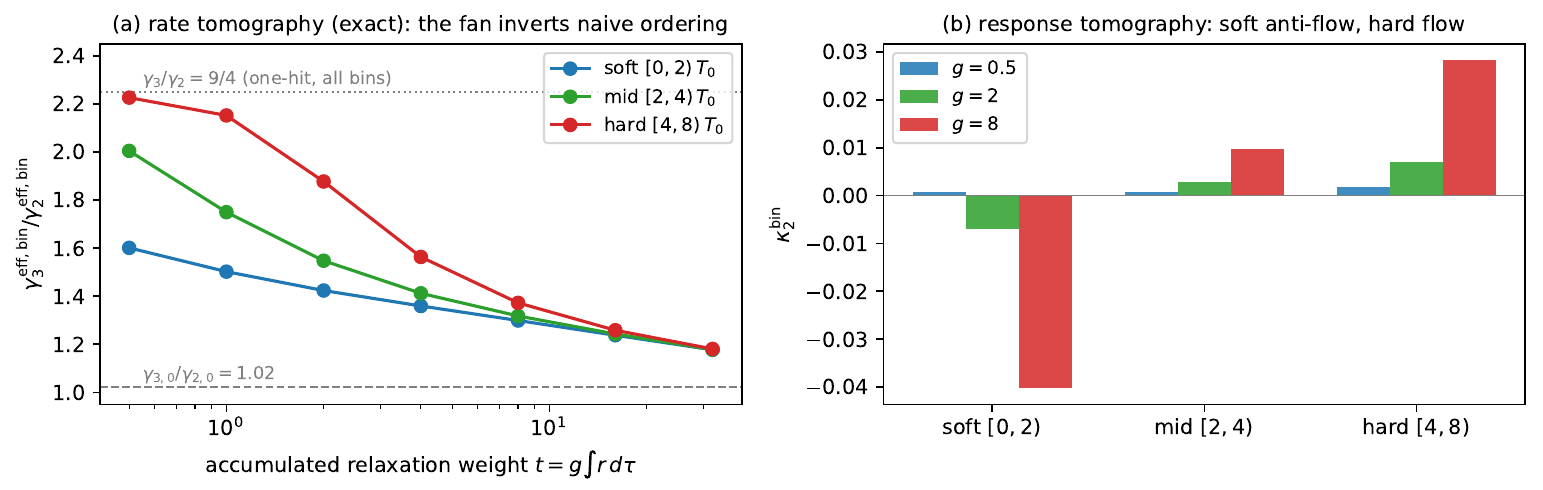}
 \caption{Momentum-space spectral tomography of the Fokker--Planck kernel.
 (a) Effective-rate ratio $\gamma_3^\eff/\gamma_2^\eff$ in three momentum bins
 versus the accumulated relaxation weight $t=g\!\int\!r\,\dd\tau$. All bins
 start at the thermal one-hit value $9/4$ and approach the late-time value
 $50/49$ of Eq.~\eqref{eq:late}; the soft bin gets there first, the hard bin
 last. (b) Momentum-differential elliptic response at $g=0.5,2,8$: flow
 migrates to hard momenta while the soft component turns anti-aligned.}
 \label{fig:tomography}
\end{figure*}

\textit{Numerical verification.---}
We solve the momentum-resolved model for $F(\xperp,\phi,p,\tau)=p^2f$ with the
same exact spectral advection as before and a collision step given by the
matrix exponential of a conservative Sturm--Liouville discretization of
$\Lambda_l$ (production grid $N_x=80$, $N_\phi=32$, $N_p=20$, $p_{\max}=12T_0$).
Energy and momentum are conserved to $10^{-14}$; seeded eigenmodes decay at
their analytic rates; the full solution reproduces the one-hit result to
better than $1\%$ at $g=0.02$ in both the direct-dominated $n=2$ and the
subtraction-dominated $n=3$ channels; and a scalar spectrum run through the
same pipeline returns $D_{32}=2.250000$. Most importantly, the direct-channel
effective eigenvalues extracted from the response, $0.9900$ and $2.2216$, agree
with the discrete-operator prediction of Eq.~\eqref{eq:effective} to seven
digits: factorization is exact channel by channel, and it is the mixture of
channels with different momentum profiles that breaks it.

Figure~\ref{fig:tomography} shows the consequence that Eq.~\eqref{eq:spectrum}
predicts. At one hit, $\Lambda_n$ acts on the thermal shape pointwise,
$\Lambda_ne^{-p/T}=(\varkappa n^2/p^2)e^{-p/T}$, so every momentum bin starts at
$9/4$ (verified to $2\times10^{-4}$). Under repeated scattering the ordering
inverts the naive expectation. One might expect the soft bin, where
$\gamma_l(p)$ is largest, to display the $l^2$ hierarchy most clearly. Instead,
a bin displays the modes that \emph{survive} in it: at soft momenta the
continuum is annihilated almost immediately, leaving the $p^{\,l}$ tails of the
near-degenerate bound states, so the soft bin reaches $50/49$ first ($1.60$ at
$t=0.5$), while the hard bin, where rates are weakest, keeps the
continuum-average value $9/4$ longest ($2.23$). The inverted fan is a
parameter-free prediction of the Rydberg structure. The response follows the
same division of labor [Fig.~\ref{fig:tomography}(b)]: the mean momentum of
the $V_2$-carrying profile rises from the thermal $3T_0$ to $5.3T_0$ by
$\tau=6R$ at $g=2$, flow is increasingly carried by hard partons, and the soft
bin counter-rotates, $\kappa_2^{\rm soft}=-4.0\times10^{-2}$ against
$\kappa_2^{\rm hard}=+2.8\times10^{-2}$ at $g=8$. A scalar kernel cannot
produce a momentum-differential sign change, because its response profile
factorizes from its angular rate.

\begin{figure*}[t]
 \includegraphics[width=0.96\textwidth]{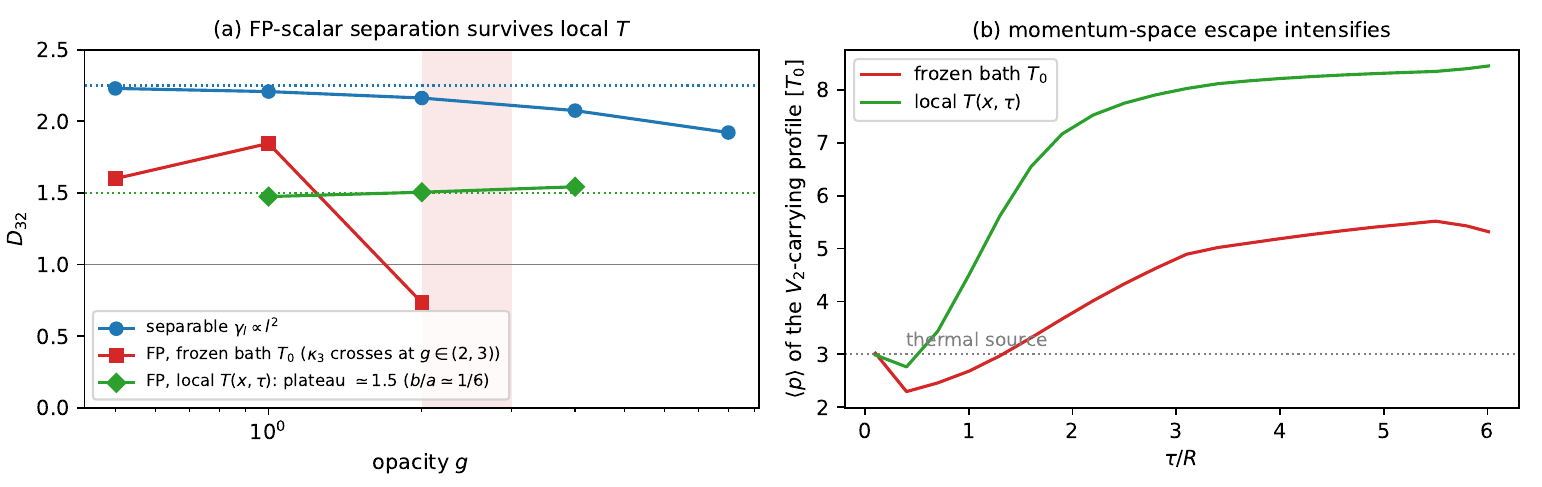}
 \caption{Robustness under a conformal local temperature $T(\xperp,\tau)$.
 (a) $D_{32}(g)$ for the separable $l^2$ spectrum, the Fokker--Planck kernel
 with frozen bath $T_0$, and the same kernel with local $T$. The local-$T$
 result forms a plateau at the mixed-spectrum value $3/2$ (dotted), separated
 from the flat baseline ($D_{32}=1$) and from angular diffusion.
 (b) Mean momentum of the $V_2$-carrying profile at $g=2$: the migration of
 flow to hard momenta survives and intensifies.}
 \label{fig:localT}
\end{figure*}

\textit{What is model dependent.}
With the bath scale frozen to $T_0$, the signed responses $\kappa_2$ and
$\kappa_3$ of the Fokker--Planck kernel cross zero at intermediate opacity
($g_3^\ast\in(2,3)$ for $\alpha=1$, and an order of magnitude earlier at the
conformal exponent $\alpha=1/3$), which no scalar spectrum does. These
crossings disappear when the operator and the Landau equilibrium are evaluated
at the conformal local temperature $T(\xperp,\tau)=T_0(e/e_{\rm ref})^{1/3}$:
the softened bath relaxes the subtraction-generated flow less, and $\kappa_3$
stays positive through $g=4$. We therefore do not claim the crossings; they
diagnose the rate model. What survives is stronger. The $\Lambda_l$ are
self-similar under $p\to pT/T_0$, so the whole spectrum simply rescales with
$T$; the hardening of the flow-carrying profile survives and intensifies
($8.5T_0$ at $\tau=6R$, Fig.~\ref{fig:localT}(b)); and the double ratio settles
on a plateau, $D_{32}=1.475,\,1.505,\,1.543$ at $g=1,2,4$
[Fig.~\ref{fig:localT}(a)], separated from the flat baseline and from angular
diffusion ($2.21$--$2.07$ over the same range). The value is not arbitrary:
$3/2$ is the mixed-spectrum ratio $(a+9b)/(a+4b)$ at $b/a=1/6$, i.e.\ the
image of a nontrivial source-weighted spectral moment, exactly as
Eq.~\eqref{eq:effective} anticipates. A scalar model with pure angular
diffusion, pure RTA, or any single rate does not reproduce it.

Finally, the $p\to0$ singularity of Eq.~\eqref{eq:localrate} is harmless: the
energy measure $p^2\dd p$ cancels it, the one-hit rates with an infrared cutoff
$p_{\min}=a$ are known in closed form, $\gamma_n^\eff(a)=\varkappa\,(n^2T+a)/(a^2+2aT+2T^2)$,
and the bound states are insensitive to a cutoff at $O((a/2T\nu)^{2l})$. Once
opacity is expressed as the accumulated physical $l=2$ relaxation weight
$\chi_2=\int\dd\tau\,\gamma_2^{\rm ref}$ rather than the raw code parameter,
all results are stable under variations of $p_{\min}$, $p_{\max}$, the inner
boundary condition, and the radial resolution (Supplemental Material).

\textit{Conclusions.}
Fundamentally, stripped of the specific heavy-ion context, this is a universal theorem of kinetic theory. In any interacting gas, the departure from equilibrium is captured by angular harmonics $l$, each relaxing at a distinct rate $\gamma_l$. While the hydrodynamic limit---and shear viscosity---is dictated solely by $\gamma_2$, a short-lived system undergoing few collisions retains memory of the entire relaxation spectrum. Its response to a geometric deformation measures this spectrum independent of the geometry itself [Eq.~\eqref{eq:ratio}]. Because the QCD collision kernel is momentum-dependent, collective flow, shear viscosity, and late-time equilibration represent three fundamentally different spectral averages [Eq.~\eqref{eq:eta}]. Consequently, two transport models tuned to the exact same viscosity can legitimately predict different flows in small systems; the discrepancy stems from hidden relaxation modes invisible to hydrodynamics.

This mathematical structure governs two-dimensional electron and atomic Fermi gases alike. There, Pauli blocking splits the spectrum by parity \cite{Ledwith:2019a}, producing scale-dependent viscosity and nonmonotonic transport \cite{Lucas:2018scl,Gurzhi:1968ur}. The factorization theorem of Eq.~\eqref{eq:ratio} applies verbatim, merely exchanging a fireball's eccentricity for a device's nonlocal conductance. The QCD kernel sits at the opposite extreme, generating a smooth quadratic hierarchy $\gamma_l\propto l^2$ rather than a parity split. In both fields, exploring intermediate-Knudsen transport is equivalent to performing spectroscopy on the microscopic collision operator \cite{Nilsson:2025odd,Maki:2025odd}.

For the heavy-ion community, these findings yield three actionable consequences:
\begin{itemize}
    \item \textbf{Transport modeling:} Viscosity matching is insufficient. Codes agreeing on $\eta/s$ but diverging in flow are not in tension; they are probing different, directly comparable relaxation spectra.
    \item \textbf{Bayesian inference:} The RTA vs EKT discrepancy is not an irreducible model error, but a parametrizable physical property that can be marginalized over or constrained.
    \item \textbf{Experiment:} Flow in small systems acquires a rigorous microscopic target: isolating the QCD relaxation spectrum, a measurement strictly impossible in the macroscopic hydrodynamic limit.
\end{itemize}

Phenomenologically, because the QCD kernel reorganizes flow in momentum space, low-$p_T$-differential $v_2$ and $v_3$ and the harmonic double ratio $D_{32}$ directly encode microscopic scattering rates. Across a system-size scan, this spectral information is progressively erased as the system approaches the hydrodynamic limit. The decomposition $A[h]+n^2B[h]$ identifies the exact parameters to promote in future global analyses \cite{Everett:2020xug,Bernhard:2019bmu}: the low-lying spectral moments of the collision operator, which must be treated on equal footing with $\eta/s$.

The deterministic solver \textsc{SolKin} utilized in this work is publicly available \cite{Soloveva:2026SolKin}.
\begin{acknowledgments}
The author thanks Andrey Sadofiev, Wilke van der Schee, Urs Wiedemann, Hendrik van Hees, D. Wagner S.~Schlichting, L.~Constantin, and C.~ Greiner for valuable
suggestions. We warmly thank Urs Wiedemann and Wilke van der Schee for their hospitality during our visit to CERN and for the insightful discussions that took place there. This work was supported by the Deutsche Forschungsgemeinschaft
through CRC-TR~211, project No.~315477589. 
We also acknowledge the use of Google Gemini for assistance with text reformulation.
\end{acknowledgments}

\bibliography{rtamodes}

\begin{thebibliography}{31}%
\makeatletter
\providecommand \@ifxundefined [1]{%
 \@ifx{#1\undefined}
}%
\providecommand \@ifnum [1]{%
 \ifnum #1\expandafter \@firstoftwo
 \else \expandafter \@secondoftwo
 \fi
}%
\providecommand \@ifx [1]{%
 \ifx #1\expandafter \@firstoftwo
 \else \expandafter \@secondoftwo
 \fi
}%
\providecommand \natexlab [1]{#1}%
\providecommand \enquote  [1]{``#1''}%
\providecommand \bibnamefont  [1]{#1}%
\providecommand \bibfnamefont [1]{#1}%
\providecommand \citenamefont [1]{#1}%
\providecommand \href@noop [0]{\@secondoftwo}%
\providecommand \href [0]{\begingroup \@sanitize@url \@href}%
\providecommand \@href[1]{\@@startlink{#1}\@@href}%
\providecommand \@@href[1]{\endgroup#1\@@endlink}%
\providecommand \@sanitize@url [0]{\catcode `\\12\catcode `\$12\catcode `\&12\catcode `\#12\catcode `\^12\catcode `\_12\catcode `\%12\relax}%
\providecommand \@@startlink[1]{}%
\providecommand \@@endlink[0]{}%
\providecommand \url  [0]{\begingroup\@sanitize@url \@url }%
\providecommand \@url [1]{\endgroup\@href {#1}{\urlprefix }}%
\providecommand \urlprefix  [0]{URL }%
\providecommand \Eprint [0]{\href }%
\providecommand \doibase [0]{https://doi.org/}%
\providecommand \selectlanguage [0]{\@gobble}%
\providecommand \bibinfo  [0]{\@secondoftwo}%
\providecommand \bibfield  [0]{\@secondoftwo}%
\providecommand \translation [1]{[#1]}%
\providecommand \BibitemOpen [0]{}%
\providecommand \bibitemStop [0]{}%
\providecommand \bibitemNoStop [0]{.\EOS\space}%
\providecommand \EOS [0]{\spacefactor3000\relax}%
\providecommand \BibitemShut  [1]{\csname bibitem#1\endcsname}%
\let\auto@bib@innerbib\@empty
\bibitem [{\citenamefont {Nagle}\ and\ \citenamefont {Zajc}(2018)}]{Nagle:2018nvi}%
  \BibitemOpen
  \bibfield  {author} {\bibinfo {author} {\bibfnamefont {J.~L.}\ \bibnamefont {Nagle}}\ and\ \bibinfo {author} {\bibfnamefont {W.~A.}\ \bibnamefont {Zajc}},\ }\bibfield  {title} {\bibinfo {title} {Small system collectivity in relativistic hadronic and nuclear collisions},\ }\href@noop {} {\bibfield  {journal} {\bibinfo  {journal} {Ann. Rev. Nucl. Part. Sci.}\ }\textbf {\bibinfo {volume} {68}},\ \bibinfo {pages} {211} (\bibinfo {year} {2018})},\ \Eprint {https://arxiv.org/abs/1801.03477} {arXiv:1801.03477 [nucl-ex]} \BibitemShut {NoStop}%
\bibitem [{\citenamefont {Schenke}(2021)}]{Schenke:2021mxx}%
  \BibitemOpen
  \bibfield  {author} {\bibinfo {author} {\bibfnamefont {B.}~\bibnamefont {Schenke}},\ }\bibfield  {title} {\bibinfo {title} {The smallest fluid on earth},\ }\href@noop {} {\bibfield  {journal} {\bibinfo  {journal} {Rept. Prog. Phys.}\ }\textbf {\bibinfo {volume} {84}},\ \bibinfo {pages} {082301} (\bibinfo {year} {2021})},\ \Eprint {https://arxiv.org/abs/2102.11189} {arXiv:2102.11189 [nucl-th]} \BibitemShut {NoStop}%
\bibitem [{\citenamefont {Brewer}\ \emph {et~al.}(2021)\citenamefont {Brewer}, \citenamefont {Mazeliauskas},\ and\ \citenamefont {van~der Schee}}]{Brewer:2021kiv}%
  \BibitemOpen
  \bibfield  {author} {\bibinfo {author} {\bibfnamefont {J.}~\bibnamefont {Brewer}}, \bibinfo {author} {\bibfnamefont {A.}~\bibnamefont {Mazeliauskas}},\ and\ \bibinfo {author} {\bibfnamefont {W.}~\bibnamefont {van~der Schee}},\ }\bibfield  {title} {\bibinfo {title} {Opportunities of oo and po collisions at the lhc},\ }\href@noop {} {\bibfield  {journal} {\bibinfo  {journal} {CERN Yellow Rep. Monogr.}\ } (\bibinfo {year} {2021})},\ \Eprint {https://arxiv.org/abs/1812.05688} {arXiv:1812.05688 [hep-ph]} \BibitemShut {NoStop}%
\bibitem [{\citenamefont {Giacalone}\ \emph {et~al.}(2025)\citenamefont {Giacalone}, \citenamefont {Bally}, \citenamefont {Nijs} \emph {et~al.}}]{Giacalone:2024luz}%
  \BibitemOpen
  \bibfield  {author} {\bibinfo {author} {\bibfnamefont {G.}~\bibnamefont {Giacalone}}, \bibinfo {author} {\bibfnamefont {B.}~\bibnamefont {Bally}}, \bibinfo {author} {\bibfnamefont {G.}~\bibnamefont {Nijs}}, \emph {et~al.},\ }\bibfield  {title} {\bibinfo {title} {The unexpected uses of a bowling pin: exploiting $^{20}$ne isotopes for precision characterizations of collectivity in small systems},\ }\href@noop {} {\bibfield  {journal} {\bibinfo  {journal} {Phys. Rev. Lett.}\ } (\bibinfo {year} {2025})},\ \Eprint {https://arxiv.org/abs/2402.05995} {arXiv:2402.05995 [nucl-th]} \BibitemShut {NoStop}%
\bibitem [{\citenamefont {Krupczak}\ \emph {et~al.}(2026)\citenamefont {Krupczak}, \citenamefont {Borghini},\ and\ \citenamefont {Roch}}]{Krupczak:2026oo}%
  \BibitemOpen
  \bibfield  {author} {\bibinfo {author} {\bibfnamefont {R.}~\bibnamefont {Krupczak}}, \bibinfo {author} {\bibfnamefont {N.}~\bibnamefont {Borghini}},\ and\ \bibinfo {author} {\bibfnamefont {H.}~\bibnamefont {Roch}},\ }\bibfield  {title} {\bibinfo {title} {Event-by-event fluctuations of elliptic flow in ultrarelativistic o+o collisions},\ }\href@noop {} {\  (\bibinfo {year} {2026})},\ \Eprint {https://arxiv.org/abs/2606.22558} {arXiv:2606.22558 [nucl-th]} \BibitemShut {NoStop}%
\bibitem [{\citenamefont {He}\ \emph {et~al.}(2016)\citenamefont {He}, \citenamefont {Edmonds}, \citenamefont {Lin}, \citenamefont {Liu}, \citenamefont {Molnar},\ and\ \citenamefont {Wang}}]{He:2015hfa}%
  \BibitemOpen
  \bibfield  {author} {\bibinfo {author} {\bibfnamefont {L.}~\bibnamefont {He}}, \bibinfo {author} {\bibfnamefont {T.}~\bibnamefont {Edmonds}}, \bibinfo {author} {\bibfnamefont {Z.-W.}\ \bibnamefont {Lin}}, \bibinfo {author} {\bibfnamefont {F.}~\bibnamefont {Liu}}, \bibinfo {author} {\bibfnamefont {D.}~\bibnamefont {Molnar}},\ and\ \bibinfo {author} {\bibfnamefont {F.}~\bibnamefont {Wang}},\ }\bibfield  {title} {\bibinfo {title} {Anisotropic parton escape is the dominant source of azimuthal anisotropy in transport models},\ }\href@noop {} {\bibfield  {journal} {\bibinfo  {journal} {Phys. Lett. B}\ }\textbf {\bibinfo {volume} {753}},\ \bibinfo {pages} {506} (\bibinfo {year} {2016})},\ \Eprint {https://arxiv.org/abs/1502.05572} {arXiv:1502.05572 [nucl-th]} \BibitemShut {NoStop}%
\bibitem [{\citenamefont {Bhatnagar}\ \emph {et~al.}(1954)\citenamefont {Bhatnagar}, \citenamefont {Gross},\ and\ \citenamefont {Krook}}]{Bhatnagar:1954zz}%
  \BibitemOpen
  \bibfield  {author} {\bibinfo {author} {\bibfnamefont {P.~L.}\ \bibnamefont {Bhatnagar}}, \bibinfo {author} {\bibfnamefont {E.~P.}\ \bibnamefont {Gross}},\ and\ \bibinfo {author} {\bibfnamefont {M.}~\bibnamefont {Krook}},\ }\bibfield  {title} {\bibinfo {title} {A model for collision processes in gases. 1. small amplitude processes in charged and neutral one-component systems},\ }\href@noop {} {\bibfield  {journal} {\bibinfo  {journal} {Phys. Rev.}\ }\textbf {\bibinfo {volume} {94}},\ \bibinfo {pages} {511} (\bibinfo {year} {1954})}\BibitemShut {NoStop}%
\bibitem [{\citenamefont {Anderson}\ and\ \citenamefont {Witting}(1974)}]{Anderson:1974nyl}%
  \BibitemOpen
  \bibfield  {author} {\bibinfo {author} {\bibfnamefont {J.~L.}\ \bibnamefont {Anderson}}\ and\ \bibinfo {author} {\bibfnamefont {H.~R.}\ \bibnamefont {Witting}},\ }\bibfield  {title} {\bibinfo {title} {A relativistic relaxation-time model for the boltzmann equation},\ }\href@noop {} {\bibfield  {journal} {\bibinfo  {journal} {Physica}\ }\textbf {\bibinfo {volume} {74}},\ \bibinfo {pages} {466} (\bibinfo {year} {1974})}\BibitemShut {NoStop}%
\bibitem [{\citenamefont {Arnold}\ \emph {et~al.}(2003)\citenamefont {Arnold}, \citenamefont {Moore},\ and\ \citenamefont {Yaffe}}]{Arnold:2002zm}%
  \BibitemOpen
  \bibfield  {author} {\bibinfo {author} {\bibfnamefont {P.~B.}\ \bibnamefont {Arnold}}, \bibinfo {author} {\bibfnamefont {G.~D.}\ \bibnamefont {Moore}},\ and\ \bibinfo {author} {\bibfnamefont {L.~G.}\ \bibnamefont {Yaffe}},\ }\bibfield  {title} {\bibinfo {title} {Effective kinetic theory for high temperature gauge theories},\ }\href@noop {} {\bibfield  {journal} {\bibinfo  {journal} {JHEP}\ }\textbf {\bibinfo {volume} {01}},\ \bibinfo {pages} {030}},\ \Eprint {https://arxiv.org/abs/hep-ph/0209353} {arXiv:hep-ph/0209353} \BibitemShut {NoStop}%
\bibitem [{\citenamefont {Kurkela}\ \emph {et~al.}(2019{\natexlab{a}})\citenamefont {Kurkela}, \citenamefont {Wiedemann},\ and\ \citenamefont {Wu}}]{Kurkela:2018qeb}%
  \BibitemOpen
  \bibfield  {author} {\bibinfo {author} {\bibfnamefont {A.}~\bibnamefont {Kurkela}}, \bibinfo {author} {\bibfnamefont {U.~A.}\ \bibnamefont {Wiedemann}},\ and\ \bibinfo {author} {\bibfnamefont {B.}~\bibnamefont {Wu}},\ }\bibfield  {title} {\bibinfo {title} {Opacity dependence of elliptic flow in kinetic theory},\ }\href@noop {} {\bibfield  {journal} {\bibinfo  {journal} {Eur. Phys. J. C}\ }\textbf {\bibinfo {volume} {79}},\ \bibinfo {pages} {759} (\bibinfo {year} {2019}{\natexlab{a}})},\ \Eprint {https://arxiv.org/abs/1805.04081} {arXiv:1805.04081 [hep-ph]} \BibitemShut {NoStop}%
\bibitem [{\citenamefont {Kurkela}\ \emph {et~al.}(2019{\natexlab{b}})\citenamefont {Kurkela}, \citenamefont {Wiedemann},\ and\ \citenamefont {Wu}}]{Kurkela:2019kip}%
  \BibitemOpen
  \bibfield  {author} {\bibinfo {author} {\bibfnamefont {A.}~\bibnamefont {Kurkela}}, \bibinfo {author} {\bibfnamefont {U.~A.}\ \bibnamefont {Wiedemann}},\ and\ \bibinfo {author} {\bibfnamefont {B.}~\bibnamefont {Wu}},\ }\bibfield  {title} {\bibinfo {title} {Flow in aa and pa as an interplay of fluid-like and non-fluid like excitations},\ }\href@noop {} {\bibfield  {journal} {\bibinfo  {journal} {Eur. Phys. J. C}\ }\textbf {\bibinfo {volume} {79}},\ \bibinfo {pages} {965} (\bibinfo {year} {2019}{\natexlab{b}})},\ \Eprint {https://arxiv.org/abs/1905.05139} {arXiv:1905.05139 [hep-ph]} \BibitemShut {NoStop}%
\bibitem [{\citenamefont {Ambru\c{s}}\ \emph {et~al.}(2022)\citenamefont {Ambru\c{s}}, \citenamefont {Schlichting},\ and\ \citenamefont {Werthmann}}]{Ambrus:2021fej}%
  \BibitemOpen
  \bibfield  {author} {\bibinfo {author} {\bibfnamefont {V.~E.}\ \bibnamefont {Ambru\c{s}}}, \bibinfo {author} {\bibfnamefont {S.}~\bibnamefont {Schlichting}},\ and\ \bibinfo {author} {\bibfnamefont {C.}~\bibnamefont {Werthmann}},\ }\bibfield  {title} {\bibinfo {title} {Development of transverse flow at small and large opacities in conformal kinetic theory},\ }\href@noop {} {\bibfield  {journal} {\bibinfo  {journal} {Phys. Rev. D}\ }\textbf {\bibinfo {volume} {105}},\ \bibinfo {pages} {014031} (\bibinfo {year} {2022})},\ \Eprint {https://arxiv.org/abs/2109.03290} {arXiv:2109.03290 [hep-ph]} \BibitemShut {NoStop}%
\bibitem [{\citenamefont {Ambru\c{s}}\ \emph {et~al.}(2023)\citenamefont {Ambru\c{s}}, \citenamefont {Schlichting},\ and\ \citenamefont {Werthmann}}]{Ambrus:2022qya}%
  \BibitemOpen
  \bibfield  {author} {\bibinfo {author} {\bibfnamefont {V.~E.}\ \bibnamefont {Ambru\c{s}}}, \bibinfo {author} {\bibfnamefont {S.}~\bibnamefont {Schlichting}},\ and\ \bibinfo {author} {\bibfnamefont {C.}~\bibnamefont {Werthmann}},\ }\bibfield  {title} {\bibinfo {title} {Establishing the range of applicability of hydrodynamics in high-energy collisions},\ }\href@noop {} {\bibfield  {journal} {\bibinfo  {journal} {Phys. Rev. Lett.}\ }\textbf {\bibinfo {volume} {130}},\ \bibinfo {pages} {152301} (\bibinfo {year} {2023})},\ \Eprint {https://arxiv.org/abs/2211.14356} {arXiv:2211.14356 [hep-ph]} \BibitemShut {NoStop}%
\bibitem [{\citenamefont {Kurkela}\ \emph {et~al.}(2019{\natexlab{c}})\citenamefont {Kurkela}, \citenamefont {Mazeliauskas}, \citenamefont {Paquet}, \citenamefont {Schlichting},\ and\ \citenamefont {Teaney}}]{Kurkela:2018vqr}%
  \BibitemOpen
  \bibfield  {author} {\bibinfo {author} {\bibfnamefont {A.}~\bibnamefont {Kurkela}}, \bibinfo {author} {\bibfnamefont {A.}~\bibnamefont {Mazeliauskas}}, \bibinfo {author} {\bibfnamefont {J.-F.}\ \bibnamefont {Paquet}}, \bibinfo {author} {\bibfnamefont {S.}~\bibnamefont {Schlichting}},\ and\ \bibinfo {author} {\bibfnamefont {D.}~\bibnamefont {Teaney}},\ }\bibfield  {title} {\bibinfo {title} {Matching the nonequilibrium initial stage of heavy ion collisions to hydrodynamics with qcd kinetic theory},\ }\href@noop {} {\bibfield  {journal} {\bibinfo  {journal} {Phys. Rev. Lett.}\ }\textbf {\bibinfo {volume} {122}},\ \bibinfo {pages} {122302} (\bibinfo {year} {2019}{\natexlab{c}})},\ \Eprint {https://arxiv.org/abs/1805.01604} {arXiv:1805.01604 [hep-ph]} \BibitemShut {NoStop}%
\bibitem [{\citenamefont {Kurkela}\ \emph {et~al.}(2019{\natexlab{d}})\citenamefont {Kurkela}, \citenamefont {Mazeliauskas}, \citenamefont {Paquet}, \citenamefont {Schlichting},\ and\ \citenamefont {Teaney}}]{Kurkela:2018wud}%
  \BibitemOpen
  \bibfield  {author} {\bibinfo {author} {\bibfnamefont {A.}~\bibnamefont {Kurkela}}, \bibinfo {author} {\bibfnamefont {A.}~\bibnamefont {Mazeliauskas}}, \bibinfo {author} {\bibfnamefont {J.-F.}\ \bibnamefont {Paquet}}, \bibinfo {author} {\bibfnamefont {S.}~\bibnamefont {Schlichting}},\ and\ \bibinfo {author} {\bibfnamefont {D.}~\bibnamefont {Teaney}},\ }\bibfield  {title} {\bibinfo {title} {Effective kinetic description of event-by-event pre-equilibrium dynamics in high-energy heavy-ion collisions},\ }\href@noop {} {\bibfield  {journal} {\bibinfo  {journal} {Phys. Rev. C}\ }\textbf {\bibinfo {volume} {99}},\ \bibinfo {pages} {034910} (\bibinfo {year} {2019}{\natexlab{d}})},\ \Eprint {https://arxiv.org/abs/1805.00961} {arXiv:1805.00961 [hep-ph]} \BibitemShut {NoStop}%
\bibitem [{\citenamefont {Bernhard}\ \emph {et~al.}(2019)\citenamefont {Bernhard}, \citenamefont {Moreland},\ and\ \citenamefont {Bass}}]{Bernhard:2019bmu}%
  \BibitemOpen
  \bibfield  {author} {\bibinfo {author} {\bibfnamefont {J.~E.}\ \bibnamefont {Bernhard}}, \bibinfo {author} {\bibfnamefont {J.~S.}\ \bibnamefont {Moreland}},\ and\ \bibinfo {author} {\bibfnamefont {S.~A.}\ \bibnamefont {Bass}},\ }\bibfield  {title} {\bibinfo {title} {Bayesian estimation of the specific shear and bulk viscosity of quark--gluon plasma},\ }\href@noop {} {\bibfield  {journal} {\bibinfo  {journal} {Nature Phys.}\ }\textbf {\bibinfo {volume} {15}},\ \bibinfo {pages} {1113} (\bibinfo {year} {2019})}\BibitemShut {NoStop}%
\bibitem [{\citenamefont {Borghini}\ \emph {et~al.}(2023)\citenamefont {Borghini}, \citenamefont {Borrell}, \citenamefont {Feld}, \citenamefont {Roch}, \citenamefont {Schlichting},\ and\ \citenamefont {Werthmann}}]{Borghini:2022vna}%
  \BibitemOpen
  \bibfield  {author} {\bibinfo {author} {\bibfnamefont {N.}~\bibnamefont {Borghini}}, \bibinfo {author} {\bibfnamefont {M.}~\bibnamefont {Borrell}}, \bibinfo {author} {\bibfnamefont {S.}~\bibnamefont {Feld}}, \bibinfo {author} {\bibfnamefont {H.}~\bibnamefont {Roch}}, \bibinfo {author} {\bibfnamefont {S.}~\bibnamefont {Schlichting}},\ and\ \bibinfo {author} {\bibfnamefont {C.}~\bibnamefont {Werthmann}},\ }\bibfield  {title} {\bibinfo {title} {Statistical analysis of initial-state and final-state response in heavy-ion collisions},\ }\href@noop {} {\bibfield  {journal} {\bibinfo  {journal} {Phys. Rev. C}\ }\textbf {\bibinfo {volume} {107}},\ \bibinfo {pages} {034905} (\bibinfo {year} {2023})},\ \Eprint {https://arxiv.org/abs/2209.01176} {arXiv:2209.01176 [nucl-th]} \BibitemShut {NoStop}%
\bibitem [{\citenamefont {Krupczak}\ \emph {et~al.}(2025)\citenamefont {Krupczak}, \citenamefont {Borghini},\ and\ \citenamefont {Roch}}]{Krupczak:2025mbm}%
  \BibitemOpen
  \bibfield  {author} {\bibinfo {author} {\bibfnamefont {R.}~\bibnamefont {Krupczak}}, \bibinfo {author} {\bibfnamefont {N.}~\bibnamefont {Borghini}},\ and\ \bibinfo {author} {\bibfnamefont {H.}~\bibnamefont {Roch}},\ }\bibfield  {title} {\bibinfo {title} {Mode-by-mode analysis of the initial state and hydrodynamic response in heavy-ion collisions in centrality classes},\ }\href@noop {} {\  (\bibinfo {year} {2025})},\ \Eprint {https://arxiv.org/abs/2508.05336} {arXiv:2508.05336 [nucl-th]} \BibitemShut {NoStop}%
\bibitem [{\citenamefont {Ledwith}\ \emph {et~al.}(2019)\citenamefont {Ledwith}, \citenamefont {Guo},\ and\ \citenamefont {Levitov}}]{Ledwith:2019a}%
  \BibitemOpen
  \bibfield  {author} {\bibinfo {author} {\bibfnamefont {P.~J.}\ \bibnamefont {Ledwith}}, \bibinfo {author} {\bibfnamefont {H.}~\bibnamefont {Guo}},\ and\ \bibinfo {author} {\bibfnamefont {L.}~\bibnamefont {Levitov}},\ }\bibfield  {title} {\bibinfo {title} {The hierarchy of excitation lifetimes in two-dimensional fermi gases},\ }\href@noop {} {\bibfield  {journal} {\bibinfo  {journal} {Annals Phys.}\ }\textbf {\bibinfo {volume} {411}},\ \bibinfo {pages} {167913} (\bibinfo {year} {2019})}\BibitemShut {NoStop}%
\bibitem [{\citenamefont {Lucas}\ and\ \citenamefont {Fong}(2018)}]{Lucas:2018scl}%
  \BibitemOpen
  \bibfield  {author} {\bibinfo {author} {\bibfnamefont {A.}~\bibnamefont {Lucas}}\ and\ \bibinfo {author} {\bibfnamefont {K.~C.}\ \bibnamefont {Fong}},\ }\bibfield  {title} {\bibinfo {title} {Hydrodynamics of electrons in graphene},\ }\href@noop {} {\bibfield  {journal} {\bibinfo  {journal} {J. Phys. Condens. Matter}\ }\textbf {\bibinfo {volume} {30}},\ \bibinfo {pages} {053001} (\bibinfo {year} {2018})},\ \Eprint {https://arxiv.org/abs/1710.08425} {arXiv:1710.08425 [cond-mat.str-el]} \BibitemShut {NoStop}%
\bibitem [{\citenamefont {Gurzhi}(1968)}]{Gurzhi:1968ur}%
  \BibitemOpen
  \bibfield  {author} {\bibinfo {author} {\bibfnamefont {R.~N.}\ \bibnamefont {Gurzhi}},\ }\bibfield  {title} {\bibinfo {title} {Hydrodynamic effects in solids at low temperature},\ }\href@noop {} {\bibfield  {journal} {\bibinfo  {journal} {Sov. Phys. Usp.}\ }\textbf {\bibinfo {volume} {11}},\ \bibinfo {pages} {255} (\bibinfo {year} {1968})}\BibitemShut {NoStop}%
\bibitem [{\citenamefont {Borghini}\ \emph {et~al.}(2018)\citenamefont {Borghini}, \citenamefont {Feld},\ and\ \citenamefont {Kersting}}]{Borghini:2018xum}%
  \BibitemOpen
  \bibfield  {author} {\bibinfo {author} {\bibfnamefont {N.}~\bibnamefont {Borghini}}, \bibinfo {author} {\bibfnamefont {S.}~\bibnamefont {Feld}},\ and\ \bibinfo {author} {\bibfnamefont {N.}~\bibnamefont {Kersting}},\ }\bibfield  {title} {\bibinfo {title} {Scaling behavior of anisotropic flow harmonics in the far-from-equilibrium regime},\ }\href@noop {} {\bibfield  {journal} {\bibinfo  {journal} {Eur. Phys. J. C}\ }\textbf {\bibinfo {volume} {78}},\ \bibinfo {pages} {832} (\bibinfo {year} {2018})},\ \Eprint {https://arxiv.org/abs/1804.05729} {arXiv:1804.05729 [nucl-th]} \BibitemShut {NoStop}%
\bibitem [{\citenamefont {Roch}\ and\ \citenamefont {Borghini}(2021)}]{Roch:2020zdl}%
  \BibitemOpen
  \bibfield  {author} {\bibinfo {author} {\bibfnamefont {H.}~\bibnamefont {Roch}}\ and\ \bibinfo {author} {\bibfnamefont {N.}~\bibnamefont {Borghini}},\ }\bibfield  {title} {\bibinfo {title} {Fluctuations of anisotropic flow from the finite number of rescatterings in a two-dimensional massless transport model},\ }\href@noop {} {\bibfield  {journal} {\bibinfo  {journal} {Eur. Phys. J. C}\ }\textbf {\bibinfo {volume} {81}},\ \bibinfo {pages} {380} (\bibinfo {year} {2021})},\ \Eprint {https://arxiv.org/abs/2012.02138} {arXiv:2012.02138 [nucl-th]} \BibitemShut {NoStop}%
\bibitem [{\citenamefont {Bachmann}\ \emph {et~al.}(2023)\citenamefont {Bachmann}, \citenamefont {Borghini}, \citenamefont {Feld},\ and\ \citenamefont {Roch}}]{Bachmann:2022cls}%
  \BibitemOpen
  \bibfield  {author} {\bibinfo {author} {\bibfnamefont {B.}~\bibnamefont {Bachmann}}, \bibinfo {author} {\bibfnamefont {N.}~\bibnamefont {Borghini}}, \bibinfo {author} {\bibfnamefont {S.}~\bibnamefont {Feld}},\ and\ \bibinfo {author} {\bibfnamefont {H.}~\bibnamefont {Roch}},\ }\bibfield  {title} {\bibinfo {title} {Even anisotropic-flow harmonics are from venus, odd ones are from mars},\ }\href@noop {} {\bibfield  {journal} {\bibinfo  {journal} {Eur. Phys. J. C}\ }\textbf {\bibinfo {volume} {83}},\ \bibinfo {pages} {114} (\bibinfo {year} {2023})},\ \Eprint {https://arxiv.org/abs/2203.13306} {arXiv:2203.13306 [nucl-th]} \BibitemShut {NoStop}%
\bibitem [{\citenamefont {Arnold}\ \emph {et~al.}(2000)\citenamefont {Arnold}, \citenamefont {Moore},\ and\ \citenamefont {Yaffe}}]{Arnold:2000dr}%
  \BibitemOpen
  \bibfield  {author} {\bibinfo {author} {\bibfnamefont {P.~B.}\ \bibnamefont {Arnold}}, \bibinfo {author} {\bibfnamefont {G.~D.}\ \bibnamefont {Moore}},\ and\ \bibinfo {author} {\bibfnamefont {L.~G.}\ \bibnamefont {Yaffe}},\ }\bibfield  {title} {\bibinfo {title} {{Transport coefficients in high temperature gauge theories. 1. Leading log results}},\ }\href {https://doi.org/10.1088/1126-6708/2000/11/001} {\bibfield  {journal} {\bibinfo  {journal} {JHEP}\ }\textbf {\bibinfo {volume} {11}},\ \bibinfo {pages} {001}},\ \Eprint {https://arxiv.org/abs/hep-ph/0010177} {arXiv:hep-ph/0010177} \BibitemShut {NoStop}%
\bibitem [{\citenamefont {Hong}\ and\ \citenamefont {Teaney}(2010)}]{Hong:2010at}%
  \BibitemOpen
  \bibfield  {author} {\bibinfo {author} {\bibfnamefont {J.}~\bibnamefont {Hong}}\ and\ \bibinfo {author} {\bibfnamefont {D.}~\bibnamefont {Teaney}},\ }\bibfield  {title} {\bibinfo {title} {{Spectral densities for hot QCD plasmas in a leading log approximation}},\ }\href {https://doi.org/10.1103/PhysRevC.82.044908} {\bibfield  {journal} {\bibinfo  {journal} {Phys. Rev. C}\ }\textbf {\bibinfo {volume} {82}},\ \bibinfo {pages} {044908} (\bibinfo {year} {2010})},\ \Eprint {https://arxiv.org/abs/1003.0699} {arXiv:1003.0699 [nucl-th]} \BibitemShut {NoStop}%
\bibitem [{\citenamefont {Gavassino}(2026)}]{Gavassino:2026zsz}%
  \BibitemOpen
  \bibfield  {author} {\bibinfo {author} {\bibfnamefont {L.}~\bibnamefont {Gavassino}},\ }\bibfield  {title} {\bibinfo {title} {{Quasinormal modes of relativistic Fokker-Planck kinetic theory}},\ }\href {https://doi.org/10.1103/6vfx-8kmm} {\bibfield  {journal} {\bibinfo  {journal} {Phys. Rev. D}\ }\textbf {\bibinfo {volume} {114}},\ \bibinfo {pages} {014018} (\bibinfo {year} {2026})},\ \Eprint {https://arxiv.org/abs/2601.19474} {arXiv:2601.19474 [nucl-th]} \BibitemShut {NoStop}%
\bibitem [{\citenamefont {Nilsson}\ \emph {et~al.}(2025)\citenamefont {Nilsson}, \citenamefont {Gran},\ and\ \citenamefont {Hofmann}}]{Nilsson:2025odd}%
  \BibitemOpen
  \bibfield  {author} {\bibinfo {author} {\bibfnamefont {E.}~\bibnamefont {Nilsson}}, \bibinfo {author} {\bibfnamefont {U.}~\bibnamefont {Gran}},\ and\ \bibinfo {author} {\bibfnamefont {J.}~\bibnamefont {Hofmann}},\ }\bibfield  {title} {\bibinfo {title} {Nonequilibrium relaxation and odd-even effect in finite-temperature electron gases},\ }\href@noop {} {\bibfield  {journal} {\bibinfo  {journal} {Phys. Rev. X}\ }\textbf {\bibinfo {volume} {15}},\ \bibinfo {pages} {041007} (\bibinfo {year} {2025})},\ \Eprint {https://arxiv.org/abs/2405.03635} {arXiv:2405.03635 [cond-mat.mes-hall]} \BibitemShut {NoStop}%
\bibitem [{\citenamefont {Maki}\ \emph {et~al.}(2025)\citenamefont {Maki}, \citenamefont {Gran},\ and\ \citenamefont {Hofmann}}]{Maki:2025odd}%
  \BibitemOpen
  \bibfield  {author} {\bibinfo {author} {\bibfnamefont {J.}~\bibnamefont {Maki}}, \bibinfo {author} {\bibfnamefont {U.}~\bibnamefont {Gran}},\ and\ \bibinfo {author} {\bibfnamefont {J.}~\bibnamefont {Hofmann}},\ }\bibfield  {title} {\bibinfo {title} {Odd-parity effect and scale-dependent viscosity in atomic quantum gases},\ }\href@noop {} {\bibfield  {journal} {\bibinfo  {journal} {Commun. Phys.}\ }\textbf {\bibinfo {volume} {8}},\ \bibinfo {pages} {319} (\bibinfo {year} {2025})},\ \Eprint {https://arxiv.org/abs/2408.02738} {arXiv:2408.02738 [cond-mat.quant-gas]} \BibitemShut {NoStop}%
\bibitem [{\citenamefont {Everett}\ \emph {et~al.}(2021)\citenamefont {Everett} \emph {et~al.}}]{Everett:2020xug}%
  \BibitemOpen
  \bibfield  {author} {\bibinfo {author} {\bibfnamefont {D.}~\bibnamefont {Everett}} \emph {et~al.} (\bibinfo {collaboration} {JETSCAPE}),\ }\bibfield  {title} {\bibinfo {title} {{Multi-system Bayesian constraints on the transport coefficients of QCD matter}},\ }\href {https://doi.org/10.1103/PhysRevC.103.054904} {\bibfield  {journal} {\bibinfo  {journal} {Phys. Rev. C}\ }\textbf {\bibinfo {volume} {103}},\ \bibinfo {pages} {054904} (\bibinfo {year} {2021})},\ \Eprint {https://arxiv.org/abs/2011.01430} {arXiv:2011.01430 [hep-ph]} \BibitemShut {NoStop}%
\bibitem [{\citenamefont {Soloveva}(2026)}]{Soloveva:2026SolKin}%
  \BibitemOpen
  \bibfield  {author} {\bibinfo {author} {\bibfnamefont {O.}~\bibnamefont {Soloveva}},\ }\href {https://doi.org/10.5281/zenodo.21498485} {\bibinfo {title} {{SolKin}}},\ \bibinfo {howpublished} {Zenodo} (\bibinfo {year} {2026}),\ \bibinfo {note} {software}\BibitemShut {NoStop}%
\end{thebibliography}%


\begin{thebibliography}{9}%
\makeatletter
\providecommand \@ifxundefined [1]{%
 \@ifx{#1\undefined}
}%
\providecommand \@ifnum [1]{%
 \ifnum #1\expandafter \@firstoftwo
 \else \expandafter \@secondoftwo
 \fi
}%
\providecommand \@ifx [1]{%
 \ifx #1\expandafter \@firstoftwo
 \else \expandafter \@secondoftwo
 \fi
}%
\providecommand \natexlab [1]{#1}%
\providecommand \enquote  [1]{``#1''}%
\providecommand \bibnamefont  [1]{#1}%
\providecommand \bibfnamefont [1]{#1}%
\providecommand \citenamefont [1]{#1}%
\providecommand \href@noop [0]{\@secondoftwo}%
\providecommand \href [0]{\begingroup \@sanitize@url \@href}%
\providecommand \@href[1]{\@@startlink{#1}\@@href}%
\providecommand \@@href[1]{\endgroup#1\@@endlink}%
\providecommand \@sanitize@url [0]{\catcode `\\12\catcode `\$12\catcode `\&12\catcode `\#12\catcode `\^12\catcode `\_12\catcode `\%12\relax}%
\providecommand \@@startlink[1]{}%
\providecommand \@@endlink[0]{}%
\providecommand \url  [0]{\begingroup\@sanitize@url \@url }%
\providecommand \@url [1]{\endgroup\@href {#1}{\urlprefix }}%
\providecommand \urlprefix  [0]{URL }%
\providecommand \Eprint [0]{\href }%
\providecommand \doibase [0]{https://doi.org/}%
\providecommand \selectlanguage [0]{\@gobble}%
\providecommand \bibinfo  [0]{\@secondoftwo}%
\providecommand \bibfield  [0]{\@secondoftwo}%
\providecommand \translation [1]{[#1]}%
\providecommand \BibitemOpen [0]{}%
\providecommand \bibitemStop [0]{}%
\providecommand \bibitemNoStop [0]{.\EOS\space}%
\providecommand \EOS [0]{\spacefactor3000\relax}%
\providecommand \BibitemShut  [1]{\csname bibitem#1\endcsname}%
\let\auto@bib@innerbib\@empty
\bibitem [{\citenamefont {Borghini}\ \emph {et~al.}(2018)\citenamefont {Borghini}, \citenamefont {Feld},\ and\ \citenamefont {Kersting}}]{Borghini:2018xum}%
  \BibitemOpen
  \bibfield  {author} {\bibinfo {author} {\bibfnamefont {N.}~\bibnamefont {Borghini}}, \bibinfo {author} {\bibfnamefont {S.}~\bibnamefont {Feld}},\ and\ \bibinfo {author} {\bibfnamefont {N.}~\bibnamefont {Kersting}},\ }\href@noop {} {\bibfield  {journal} {\bibinfo  {journal} {Eur. Phys. J. C}\ }\textbf {\bibinfo {volume} {78}},\ \bibinfo {pages} {832} (\bibinfo {year} {2018})},\ \Eprint {https://arxiv.org/abs/1804.05729} {arXiv:1804.05729 [nucl-th]} \BibitemShut {NoStop}%
\bibitem [{\citenamefont {Roch}\ and\ \citenamefont {Borghini}(2021)}]{Roch:2020zdl}%
  \BibitemOpen
  \bibfield  {author} {\bibinfo {author} {\bibfnamefont {H.}~\bibnamefont {Roch}}\ and\ \bibinfo {author} {\bibfnamefont {N.}~\bibnamefont {Borghini}},\ }\href@noop {} {\bibfield  {journal} {\bibinfo  {journal} {Eur. Phys. J. C}\ }\textbf {\bibinfo {volume} {81}},\ \bibinfo {pages} {380} (\bibinfo {year} {2021})},\ \Eprint {https://arxiv.org/abs/2012.02138} {arXiv:2012.02138 [nucl-th]} \BibitemShut {NoStop}%
\bibitem [{\citenamefont {Bachmann}\ \emph {et~al.}(2023)\citenamefont {Bachmann}, \citenamefont {Borghini}, \citenamefont {Feld},\ and\ \citenamefont {Roch}}]{Bachmann:2022cls}%
  \BibitemOpen
  \bibfield  {author} {\bibinfo {author} {\bibfnamefont {B.}~\bibnamefont {Bachmann}}, \bibinfo {author} {\bibfnamefont {N.}~\bibnamefont {Borghini}}, \bibinfo {author} {\bibfnamefont {S.}~\bibnamefont {Feld}},\ and\ \bibinfo {author} {\bibfnamefont {H.}~\bibnamefont {Roch}},\ }\href@noop {} {\bibfield  {journal} {\bibinfo  {journal} {Eur. Phys. J. C}\ }\textbf {\bibinfo {volume} {83}},\ \bibinfo {pages} {114} (\bibinfo {year} {2023})},\ \Eprint {https://arxiv.org/abs/2203.13306} {arXiv:2203.13306 [nucl-th]} \BibitemShut {NoStop}%
\bibitem [{\citenamefont {Romatschke}\ and\ \citenamefont {Strickland}(2003)}]{Romatschke:2003ms}%
  \BibitemOpen
  \bibfield  {author} {\bibinfo {author} {\bibfnamefont {P.}~\bibnamefont {Romatschke}}\ and\ \bibinfo {author} {\bibfnamefont {M.}~\bibnamefont {Strickland}},\ }\href@noop {} {\bibfield  {journal} {\bibinfo  {journal} {Phys. Rev. D}\ }\textbf {\bibinfo {volume} {68}},\ \bibinfo {pages} {036004} (\bibinfo {year} {2003})},\ \Eprint {https://arxiv.org/abs/hep-ph/0304092} {arXiv:hep-ph/0304092} \BibitemShut {NoStop}%
\bibitem [{\citenamefont {Baier}\ \emph {et~al.}(2001)\citenamefont {Baier}, \citenamefont {Mueller}, \citenamefont {Schiff},\ and\ \citenamefont {Son}}]{Baier:2000sb}%
  \BibitemOpen
  \bibfield  {author} {\bibinfo {author} {\bibfnamefont {R.}~\bibnamefont {Baier}}, \bibinfo {author} {\bibfnamefont {A.~H.}\ \bibnamefont {Mueller}}, \bibinfo {author} {\bibfnamefont {D.}~\bibnamefont {Schiff}},\ and\ \bibinfo {author} {\bibfnamefont {D.~T.}\ \bibnamefont {Son}},\ }\href@noop {} {\bibfield  {journal} {\bibinfo  {journal} {Phys. Lett. B}\ }\textbf {\bibinfo {volume} {502}},\ \bibinfo {pages} {51} (\bibinfo {year} {2001})},\ \Eprint {https://arxiv.org/abs/hep-ph/0009237} {arXiv:hep-ph/0009237} \BibitemShut {NoStop}%
\bibitem [{\citenamefont {Vogel}\ and\ \citenamefont {Risken}(1989)}]{Vogel:1989zz}%
  \BibitemOpen
  \bibfield  {author} {\bibinfo {author} {\bibfnamefont {K.}~\bibnamefont {Vogel}}\ and\ \bibinfo {author} {\bibfnamefont {H.}~\bibnamefont {Risken}},\ }\href {https://doi.org/10.1103/PhysRevA.40.2847} {\bibfield  {journal} {\bibinfo  {journal} {Phys. Rev. A}\ }\textbf {\bibinfo {volume} {40}},\ \bibinfo {pages} {2847} (\bibinfo {year} {1989})}\BibitemShut {NoStop}%
\bibitem [{\citenamefont {Gavassino}(2026{\natexlab{a}})}]{Gavassino:2026zsz}%
  \BibitemOpen
  \bibfield  {author} {\bibinfo {author} {\bibfnamefont {L.}~\bibnamefont {Gavassino}},\ }\href {https://doi.org/10.1103/6vfx-8kmm} {\bibfield  {journal} {\bibinfo  {journal} {Phys. Rev. D}\ }\textbf {\bibinfo {volume} {114}},\ \bibinfo {pages} {014018} (\bibinfo {year} {2026}{\natexlab{a}})},\ \Eprint {https://arxiv.org/abs/2601.19474} {arXiv:2601.19474 [nucl-th]} \BibitemShut {NoStop}%
\bibitem [{\citenamefont {Gavassino}(2026{\natexlab{b}})}]{Gavassino:2026tvy}%
  \BibitemOpen
  \bibfield  {author} {\bibinfo {author} {\bibfnamefont {L.}~\bibnamefont {Gavassino}},\ }\href {https://doi.org/10.1103/rkz9-xps2} {\bibfield  {journal} {\bibinfo  {journal} {Phys. Rev. Lett.}\ }\textbf {\bibinfo {volume} {137}},\ \bibinfo {pages} {022302} (\bibinfo {year} {2026}{\natexlab{b}})},\ \Eprint {https://arxiv.org/abs/2601.19464} {arXiv:2601.19464 [gr-qc]} \BibitemShut {NoStop}%
\bibitem [{\citenamefont {Soloveva}(2026)}]{Soloveva:2026SolKin}%
  \BibitemOpen
  \bibfield  {author} {\bibinfo {author} {\bibfnamefont {O.}~\bibnamefont {Soloveva}},\ }\href {https://doi.org/10.5281/zenodo.21498485} {\bibinfo {title} {{SolKin}}},\ \bibinfo {howpublished} {Zenodo} (\bibinfo {year} {2026}),\ \bibinfo {note} {software}\BibitemShut {NoStop}%
\end{thebibliography}%
\end{document}


\title{Supplemental Material for\\ ``Beyond Viscosity Matching: Microscopic Relaxation in Anisotropic Flow''}

\author{Olga Soloveva}
\email{soloveva@itp.uni-frankfurt.de}
\affiliation{GSI Helmholtzzentrum f\"ur Schwerionenforschung GmbH, Planckstra\ss e 1, 64291 Darmstadt, Germany}
\affiliation{Institut f\"ur Theoretische Physik, Johann Wolfgang Goethe-Universit\"at, Max-von-Laue-Str.~1, 60438 Frankfurt am Main, Germany}
\affiliation{Helmholtz Research Academy Hessen for FAIR (HFHF), GSI Helmholtz Center for Heavy Ion Physics, Campus Frankfurt, 60438 Frankfurt am Main, Germany}

\date{\today}

\maketitle

\section{2D scalar theory}
\label{sec:2d}

\subsection{Conventions and collision operator}
\label{sec:setup}
We consider a gas of massless partons confined to the transverse plane, the
two-dimensional model used in the Bielefeld transport literature to study the
onset of anisotropic flow \cite{Borghini:2018xum,Roch:2020zdl,Bachmann:2022cls}.
The phase-space density is $f(\xperp,\pvec,\tau)$, and each parton moves with
unit speed in the direction $\vperp=\pvec/p=(\cos\phi,\sin\phi)$. Since the
collision kernels of this section act only on the momentum angle $\phi$, the
modulus $p$ can be integrated out. We do so with the energy weight, which is
the weight that makes the resulting quantity a conserved density,
\begin{equation}
  \Phi(\xperp,\phi,\tau)\equiv\int_0^\infty\dd p\,p^2f(\xperp,\pvec,\tau),\qquad
  e(\xperp,\tau)=\int_0^{2\pi}\dd\phi\,\Phi ,
  \label{eq:Phi-def}
\end{equation}
so that $\Phi$ is the energy density per unit momentum angle and $e$ the
energy density. Its angular structure is resolved in circular harmonics,
\begin{equation}
  \Phi(\xperp,\phi,\tau)=\sum_{l=-\infty}^{\infty}\Phi_l(\xperp,\tau)\,e^{il\phi},
  \qquad
  \Phi_l=\frac1{2\pi}\int_0^{2\pi}\dd\phi\,e^{-il\phi}\,\Phi ,
  \label{eq:harmonics}
\end{equation}
where $\Phi_0$ carries the energy density, $\Phi_{\pm1}$ the momentum
density, and $\Phi_{\pm2}$ the shear stress; the higher harmonics are the
non-hydrodynamic content of the distribution. Because the partons are
massless and the kernel angle-only, $\Phi$ obeys a closed Boltzmann equation,
\begin{equation}
  \bigl(\partial_\tau+\vperp\cdot\nabla_\perp\bigr)\Phi=C[\Phi].
  \label{eq:boltzmann}
\end{equation}
nisotropic flow is the angular structure of the momentum distribution
integrated over the whole system. We define the energy-weighted flow
coefficients
\begin{equation}
  V_n(\tau)=\frac{\int\dd^2x\int\dd\phi\,e^{in\phi}\,\Phi}
                 {\int\dd^2x\int\dd\phi\,\Phi}
  =\frac{2\pi}{\Etot}\int\dd^2x\,\Phi_{-n}(\xperp,\tau),
  \label{eq:vn-def}
\end{equation}
with $\Etot$ the total energy, and denote their asymptotic magnitudes by
$v_n=|V_n(\tau\to\infty)|$. The eccentricity-to-flow response
\begin{equation}
  \kappa_n\equiv\frac{v_n}{\varepsilon_n}
  \label{eq:kappa-def}
\end{equation}
measures how efficiently the $n$th spatial eccentricity $\varepsilon_n$ of
the initial energy profile is converted into the $n$th momentum-space
harmonic; it is the object on which every result below is stated.

The collision operator is linearized around the Landau-matched local
equilibrium $\Phi^{\eq}$. Any kernel that is rotation invariant in the local
rest frame is diagonal in the circular harmonics, so the most general
linearized scalar model is \meq{1}: each harmonic $l$ relaxes toward its
equilibrium value at its own rate $\gamma_l(\xperp,\tau)$. The two lowest
rates are not free. Energy conservation requires $\int\dd\phi\,C=0$ and
momentum conservation $\int\dd\phi\,\vperp C=0$, and since $\vperp$ has
harmonic content $l=\pm1$ only, these conditions are equivalent to
\begin{equation}
  \gamma_0=\gamma_{\pm1}=0 .
  \label{eq:conservation}
\end{equation}
An angular-diffusion operator $D\partial_\phi^2$, for instance, has
$\gamma_1=D\neq0$ and violates momentum conservation; the spectra below are
all corrected for this. For $l\ge2$ we consider the flat relaxation-time
approximation, $\gamma_l=1/\tau_R$, in which all non-hydrodynamic harmonics
decay at one rate; angular diffusion, $\gamma_l=Dl^2$, the small-angle limit
in which the rate grows quadratically with the harmonic order
($\gamma_3/\gamma_2=9/4$); its momentum-conserving version,
$\gamma_l=D(l^2-1)$ ($\gamma_3/\gamma_2=8/3$); and the mixed spectrum that
interpolates between a large-angle and a small-angle component,

\begin{equation}
  \gamma_l=a+bl^2,\qquad\frac{\gamma_3}{\gamma_2}=\frac{a+9b}{a+4b}\in\bigl[1,\tfrac94\bigr].
  \label{eq:mixed}
\end{equation}
The ratio $b/a$ is the parameter that the single-$\tau_R$ approximation sets
to zero. Throughout Secs.~\ref{sec:2d}--\ref{sec:threed} the rates are taken
to be separable,
\begin{equation}
  \gamma_l(\xperp,\tau)=\hat g_l\,W(\xperp,\tau),
  \label{eq:locality}
\end{equation}
with one spatial profile $W$ shared by all harmonics and the spectrum
carried entirely by the constants $\hat g_l$; we use $W=e^\alpha$ with
$\alpha=1$ for a rate proportional to the density and $\alpha=1/3$ for a
conformal rate proportional to the temperature. This is the single structural
assumption of the scalar theory, and Sec.~\ref{sec:fp} is devoted to the kernel that violates it. Projecting Eq.~\eqref{eq:boltzmann} onto harmonic
$m$, using $\cos\phi\,e^{im\phi}=\tfrac12(e^{i(m+1)\phi}+e^{i(m-1)\phi})$ and
its $\sin\phi$ analogue, gives the exact ladder
\begin{equation}
  \partial_\tau\Phi_m+\tfrac12(\partial_x-i\partial_y)\Phi_{m-1}
  +\tfrac12(\partial_x+i\partial_y)\Phi_{m+1}= \label{eq:ladder}
\end{equation}
$$ = -\gamma_m\bigl[\Phi_m-\Phi^{\eq}_m\bigr]:$$
the drift couples adjacent harmonics only, and the collision term is diagonal.

\subsection{One-hit expansion and factorization theorem}
\label{sec:onehit}
Expanding in the number of collisions, $\Phi=\Phi^{(0)}+\Phi^{(1)}+\dots$, with $\Phi^{(0)}(\xperp,\phi,\tau)=\Phi_0(\xperp-\vperp t,\phi)$, $t=\tau-\tau_0$, and an isotropic initial condition $\Phi_0=E_0(\xperp)/2\pi$:

\begin{lemma}[Free streaming generates no flow]
\label{prop:nofsflow}
$\int\dd^2x\,\Phi^{(0)}(\xperp,\phi,\tau)=\Etot/2\pi$ for all $\phi$ and $\tau$, hence $V_n^{(0)}=0$ for all $n\ge1$.
\end{lemma}
\begin{proof}
The spatial integral at fixed $\phi$ is invariant under $\xperp\to\xperp-\vperp t$, and $\Phi_0$ is isotropic.
\end{proof}

\begin{lemma}[One-hit formula]
\label{prop:master}
Inserting $\Phi^{(1)}=\int_{\tau_0}^{\tau}\dd\tau'\,C[\Phi^{(0)}](\xperp-\vperp(\tau-\tau'),\phi,\tau')$ into Eq.~\eqref{eq:vn-def}, shifting $\xperp\to\xperp+\vperp(\tau-\tau')$ (unit Jacobian), and using the diagonality of $C$,
$$V_n^{(1)}=-\hat g_n\,G_n,\qquad$$
\begin{equation}
  G_n\equiv\frac{2\pi}{\Etot}\int_{\tau_0}^{\infty}\!\dd\tau\!\int\!\dd^2x\,W\bigl[\Phi^{(0)}_{-n}-\Phi^{\eq[0]}_{-n}\bigr],
  \label{eq:master}
\end{equation}
where $\Phi^{\eq[0]}$ is the equilibrium Landau-matched to the free-streaming background.
\end{lemma}
The geometry functional $G_n$ depends on $E_0$ and $W$ but not on $\{\hat g_l\}$. The equilibrium subtraction is not small (free streaming builds up local anisotropies of order one at $t\sim R$), but it enters with the same prefactor $\hat g_n$.

\begin{theorem}[Ratio theorem; \meq{4}]
\label{thm:ratio}
At leading order in opacity, independently of $E_0$, of $\alpha$, and of the equilibrium subtraction,
$(\kappa_n/\kappa_m)|_{\{\gamma_l\}}\big/(\kappa_n/\kappa_m)|_{\rm flat}=\gamma_n/\gamma_m$.
\end{theorem}
\begin{proof}
From Eq.~\eqref{eq:master}, $\kappa_n=\hat g_n|G_n|/\varepsilon_n$; the factors $|G_n|/\varepsilon_n$ cancel in the double ratio.
\end{proof}
We write $D_{nm}\equiv(\kappa_n/\kappa_m)|_{\rm spec}/(\kappa_n/\kappa_m)|_{\rm flat}$; the dilute anchors are $D_{32}=9/4,\,8/3$ and $D_{42}=4,\,5$ for the two diffusion spectra.

\subsection{Explicit geometry factor, selection rule, and regulator}
\label{sec:explicit}

With $E_0(\xperp)=\int\frac{\dd^2k}{(2\pi)^2}\tilde E(\kvec)e^{i\kvec\cdot\xperp}$ and $\int_0^{2\pi}\frac{\dd\phi}{2\pi}e^{in\phi}e^{-ikt\cos(\phi-\theta_k)}=(-i)^nJ_n(kt)e^{in\theta_k}$, the direct term for $\alpha=1$ is
\begin{equation}
  \int\dd^2x\,e^{(0)}\Phi^{(0)}_{-n}
  =\frac{(-i)^n}{2\pi}\int\frac{\dd^2k}{(2\pi)^2}\bigl|\tilde E(\kvec)\bigr|^2J_0(kt)J_n(kt)e^{in\theta_k}.
  \label{eq:powerspectrum}
\end{equation}
For the eccentric Gaussian $E_0=\frac{\Etot}{2\pi R^2}e^{-r^2/2R^2}[1+\beta_n(r/R)^n\cos n\theta]$, with $\varepsilon_n=-\beta_n2^{n/2-1}n!/\Gamma(\tfrac n2+1)$, one has $\tilde E(\kvec)=\Etot e^{-k^2R^2/2}[1+\beta_n(-i)^n(kR)^n\cos n\theta_k]$. To linear order in $\beta_n$, $|\tilde E|^2\simeq\tilde{\bar E}^2+2\tilde{\bar E}\,\mathrm{Re}[\delta\tilde E]$ with $\delta\tilde E\propto(-i)^n$, real for even and imaginary for odd $n$:

\begin{corollary}[Even/odd selection rule]
\label{prop:evenodd}
For $\alpha=1$ the direct one-hit response of all odd harmonics vanishes at linear order in the eccentricity.
\end{corollary}

For $n=1$ this reproduces translation invariance. For $n\ge3$ odd response is generated by $\alpha\neq1$ and by the equilibrium subtraction, both with prefactor $\hat g_n$, so Theorem~\ref{thm:ratio} is unaffected.

For even $n$ the geometry factor is
\begin{equation}
  v_n=\hat g_n\frac{\Etot|\beta_n|}{2\pi}\mathcal J_n,\qquad
  \end{equation}
  
  \begin{equation}
  \mathcal J_n=\frac1R\int_0^\infty\!\dd s\,\frac{s_0}{s_0+s}\int_0^\infty\!\dd u\,u^{n+1}e^{-u^2}J_0(us)J_n(us)
  \label{eq:Jn-reg}
\end{equation}
$$\simeq\frac{\Gamma(\tfrac{n+1}2)}{2\pi}\ln\frac{cR}{\tau_0}+\text{finite},$$
$u=kR$, $s=t/R$. Without the factor $s_0/(s_0+s)$ the $s$ integral diverges logarithmically for even $n$, since $J_0J_n\to\cos(n\pi/2)/\pi z$ and the free-streaming density along a ray dilutes only as $1/t$; longitudinal boost-invariant expansion supplies the $\tau_0/\tau$ dilution that regulates it. This factor modifies the background attenuation but not the ladder \eqref{eq:ladder}, and is derived rather than modeled in Sec.~\ref{sec:threed}.

\section{Beyond one collision}
\label{sec:beyond}

\subsection{Resummation and Chapman--Enskog limit}
\label{sec:CE}

Keeping the collision term in the $l=n$ channel to all orders, $\Phi_n$ obeys $\dd\Phi_n/\dd\tau=S_n-\gamma_n\Phi_n$ along a characteristic, so
\begin{equation}
  \Phi_n(\infty)=\int\dd\tau\,S_n(\tau)\exp\Bigl[-\int_\tau^\infty\gamma_n\dd\tau'\Bigr]:
  \label{eq:resummation}
\end{equation}
the rate that generates $v_n$ at low opacity attenuates it at high opacity (Gurzhi-type crossover). At first order in gradients the ladder \eqref{eq:ladder} closes on $\gamma_2\delta\Phi_{\pm2}=-\tfrac12(\partial_x\mp i\partial_y)\Phi_{\pm1}$, giving $\eta\propto(e+P)/\gamma_2$, $\tau_\pi=1/\gamma_2$ [\meq{5}]; $\gamma_{l\ge3}$ first enter at third order in gradients.

\subsection{Kinetic solver and extended-opacity scan}
\label{sec:method}

The scalar model is solved with a deterministic scheme: exact spectral translation per momentum-angle slice (Lemma S1~\ref{prop:nofsflow} holds identically) and the exact per-harmonic exponential relaxation toward the closed-form Landau-matched equilibrium, stable at any opacity and conserving energy and momentum to $10^{-14}$. The initial state $E_0\propto e^{-r^2/2R^2}[1+\tfrac\beta2(r/R)^n\cos n\theta]^2$ is non-negative and agrees with the eccentric Gaussian at linear order; $\beta=0.04,\,0.01,\,0.002$ for $n=2,3,4$ keeps the response linear to $<1\%$, and the $\beta^2$ term is removed by centered differencing. The opacity parameter $g$ multiplies $r(\xperp,\tau)=g(e/e_{\rm ref})^\alpha(\tau_0/\tau)$ with $\hat g_2=1$. Validation: dilute double ratios reproduce the anchors to four digits ($2.2500$, $2.6667$; $4.0000$, $5.0000$); the full solution agrees with Eq.~\eqref{eq:master} to $<1\%$ for $g\le0.03$; $\kappa_2$ is spectrum independent over the full opacity range for spectra normalized to the same $\hat g_2$, so that all spectrum dependence resides in $n\ge3$.

\begin{figure*}[t]
\centering
\includegraphics[width=\linewidth]{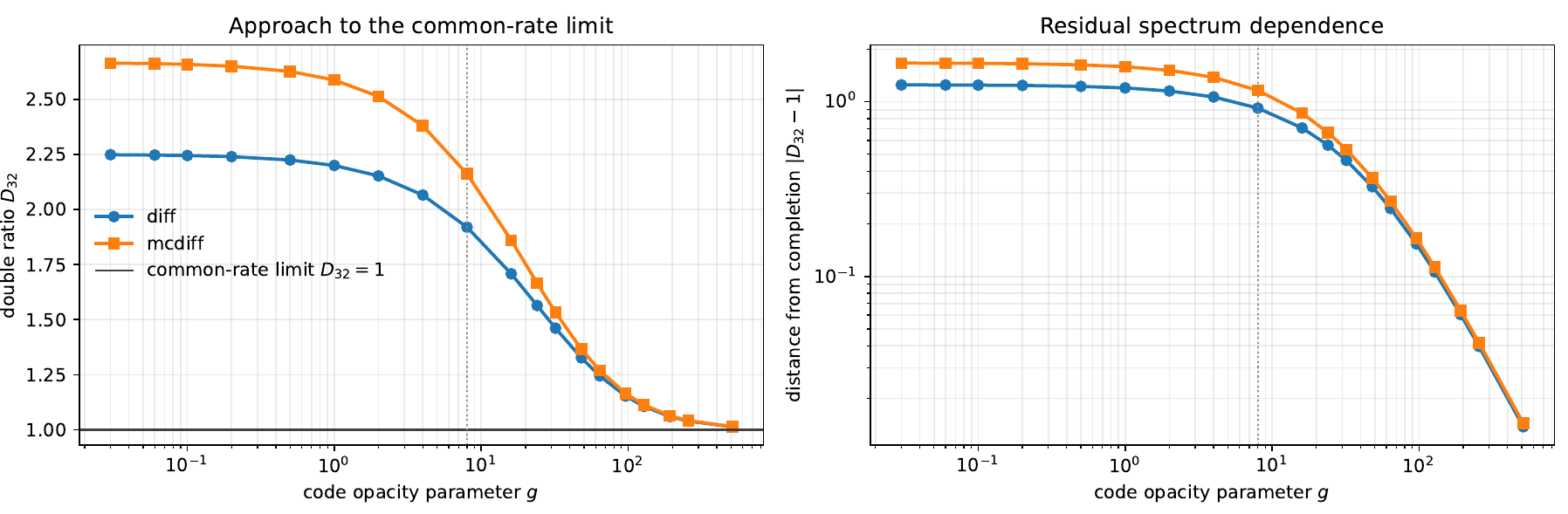}
\caption{Extended-opacity evolution of $D_{32}$ for $\gamma_l\propto l^2$ (diff) and $\gamma_l\propto l^2-1$ (mcdiff). Left: $D_{32}(g)$ from the dilute anchors ($9/4$, $8/3$) to the common-rate limit $D_{32}=1$; dotted line at $g=8$. Right: $|D_{32}-1|$; both spectra follow a common power-law asymptote (local exponent $\simeq-1.35$) and reach $|D_{32}-1|<0.1$ only for $g\gtrsim130$.}
\label{fig:d32ext}
\end{figure*}

Figure~\ref{fig:d32ext} shows $D_{32}$ over four decades. For $g\le0.1$ it is flat at $2.2467$ and $2.6615$ (within $0.2\%$ of the anchors); it decreases to $1.92$ and $2.16$ at $g=8$ and approaches $1$ as a power of the inverse opacity: $|D_{32}-1|<0.5$ for $g\gtrsim30$, $<0.1$ for $g\gtrsim1.3\times10^2$, $<0.05$ for $g\gtrsim2.2\times10^2$. At $g=8$, $70$--$74\%$ of the anchor-to-unity span survives.

\subsection{Signed quartic response}
\label{sec:kappa4}

In the dilute limit $\kappa_4<0$: the collisional $v_4$ is anti-aligned with the $\varepsilon_4$ participant plane (Fig.~\ref{fig:kappa4}), with $\kappa_4^{\rm diff}/\kappa_4^{\rm flat}=4.0000$ and $\kappa_4^{\rm mcdiff}/\kappa_4^{\rm flat}=5.0000$ at $g=0.03$. The sign is stable under $N_\phi=64\to128$ (change $10^{-8}$), initial-condition family, and amplitude. Because the even-$n$ geometry integral carries the $\ln(R/\tau_0)$ of Eq.~\eqref{eq:Jn-reg}, the $\tau_0$ dependence was tested at matched elliptic opacity [$g(\tau_0)$ adjusted so that the one-hit flat $\kappa_2$ is fixed]: $\kappa_4$ remains negative for all spectra over $0.025\le\tau_0/R\le0.20$ with a residual variation of $\pm11\%$. With increasing opacity $|\kappa_4|$ first grows, reaches a minimum at $g\simeq2$--$4$, and crosses zero at a spectrum-dependent $g_4^*$ (later for steeper spectra). At leading order $\kappa_4=\hat g_4G_4$ has a spectrum-independent sign, so the crossing and its ordering are next-to-leading-order observables not normalized to any baseline; near $g_4^*$, $D_{42}$ is ill-conditioned.

\section{3D kinematics: the eigenvalue tower and longitudinal squeeze}
\label{sec:threed}

\subsection{Kinetic equation and kernel on the sphere}
\label{sec:threed-kinematics}

With $\vhat=(\sin\theta_p\cos\phi_p,\sin\theta_p\sin\phi_p,\cos\theta_p)$, $u\equiv\cos\theta_p$, and $\Phi\equiv\int_0^\infty\dd p\,p^3f$, carrying the Bjorken term $-(p_z/\tau)\partial_{p_z}$ through the $p$ integration [$p_z\partial_{p_z}|_{p_T}=pu^2\partial_p+u(1-u^2)\partial_u$, $\int p^4\partial_pf\,\dd p=-4\Phi$] gives, at midrapidity,
\begin{equation}
  \partial_\tau\Phi+\vperp\cdot\nabla_\perp\Phi
  +\frac1\tau\Bigl[4u^2\Phi-u(1-u^2)\partial_u\Phi\Bigr]=C[\Phi],\label{eq:boltzmann3d}
\end{equation}
$$\qquad\vperp=\sqrt{1-u^2}(\cos\phi_p,\sin\phi_p).$$
The bracket is the longitudinal redshift; integrating over $\Omega$ reproduces $\partial_\tau e+\nabla_\perp\cdot\mathbf T^{\tau\perp}=-(e+P_L)/\tau$ with $P_L=\int\dd\Omega\,u^2\Phi$. In the basis $\Phi=\sum_{l,m}\Phi_{lm}\Ylm lm$, a rotation-invariant linearized operator is, by Schur's lemma, diagonal in $l$ and degenerate in $m$,
\begin{equation}
  C[\Phi]=-\sum_{l,m}\gamma_l\bigl[\Phi_{lm}-\Phi^{\eq}_{lm}\bigr]\Ylm lm,\qquad\gamma_0=\gamma_1=0 .
  \label{eq:kernel3d}
\end{equation}
The eigenvalue label is $l$, whereas the flow harmonic $n$ is the azimuthal label $m$; in two dimensions these coincide. Diffusion on $S^2$ has $\gamma_l=Dl(l+1)$, which violates momentum conservation at $l=1$; the corrected spectrum $\gamma_l=D[l(l+1)-2]$ vanishes at $l=1$ automatically. Table~\ref{tab:anchors}  lists the dilute anchors; the momentum-conserving 3D values lie inside the 2D range.

\begin{table}[t]
\centering
\caption{Dilute-limit anchor ratios in two and three dimensions.}
\label{tab:anchors}
\begin{tabular}{lcccc}
\toprule
 & \multicolumn{2}{c}{2D} & \multicolumn{2}{c}{3D} \\
 \cmidrule(lr){2-3}\cmidrule(lr){4-5}
 & $l^2$ & $l^2-1$ & $l(l{+}1)$ & $l(l{+}1)-2$ \\
\midrule
$\gamma_3/\gamma_2$ & $9/4$ & $8/3$ & $2$    & $5/2$ \\
$\gamma_4/\gamma_2$ & $4$   & $5$   & $10/3$ & $9/2$ \\
\bottomrule
\end{tabular}
\end{table}

\subsection{Parity tower and weighted ratio theorem}
\label{sec:threed-parity}

Boost invariance at midrapidity imposes $p_z\to-p_z$, under which $\Ylm lm\to(-1)^{l+m}\Ylm lm$, so only $l+m$ even survives. The measured $v_n$ is the $e^{in\phi_p}$ moment of the $u=0$ slice, whose kernel has weights $\propto\Ylm ln(\pi/2,0)$, non-zero on the tower
\begin{equation}
  \{\Ylm ln:l\ge n,\;l+n\ \text{even}\}=\{\Ylm nn,\Ylm{n+2}n,\Ylm{n+4}n,\dots\}.
  \label{eq:tower}
\end{equation}
The drift couples $(l,m)\to(l\pm1,m\pm1)$ and the redshift $\Delta l=0,\pm2$ at fixed $m$; the tower members above $(n,n)$ are populated at $\mathcal O(\varepsilon_n)$ with $\mathcal O(1)$ coefficients by $t\sim R$. The one-hit expansion proceeds as in Sec.~\ref{sec:onehit} [free-streaming background of Romatschke--Strickland type\cite{Romatschke:2003ms}; Landau-matched equilibrium $\Phi^{\eq}=e_{\rm LRF}/4\pi(u^\mu\hat v_\mu)^4$, whose $T^{\mu\nu}$ is the ideal conformal tensor], and becomes a tower sum $V_n=-\mathcal N\sum_{l\ge n,\,l+n\,{\rm even}}c_l^{(n)}\hat g_l\int\dd\tau\,\dd^2x\,W[\Phi^{(0)}-\Phi^{\eq[0]}]_{l,m=n}$ with kernel-independent $c_l^{(n)}\propto\Ylm ln(\pi/2,0)$.

\begin{theorem}[Weighted ratio theorem]
\label{thm:ratio3d}
For a single tower member the factorization is exact and the double ratio equals $\gamma_l/\gamma_{l'}$. For the physical $v_n$,
\begin{equation}
  \kappa_n=\sum_lw_{nl}\hat g_l,\qquad
  \frac{(\kappa_3/\kappa_2)|_{\{\gamma_l\}}}{(\kappa_3/\kappa_2)|_{\rm flat}}
  =\frac{\langle\gamma\rangle_{3\text{-tower}}}{\langle\gamma\rangle_{2\text{-tower}}},
  \label{eq:ratiotheorem3d}
\end{equation}
with kernel-independent geometry weights $w_{nl}$.
\end{theorem}

\subsection{Tower weights in the squeezed limit}
\label{sec:threed-weights}

Along the characteristics $p_z\tau=\mathrm{const}$ an initially isotropic distribution acquires longitudinal width $\Delta(\tau)\simeq\tau_0/\tau$, and the sheet amplitude $S=\int\dd u\,\Phi$ obeys the 2D free-streaming dynamics exactly. For $\Phi=S(\phi)\delta_\Delta(u)$, $\Phi_{ln}=2\pi S_n\Ylm ln(\pi/2,0)F_l^{(n)}(\Delta)$ with
\begin{equation}
  \bigl|\Ylm ln(\pi/2,0)\bigr|^2=\frac{2l+1}{4\pi}\frac{(l+n-1)!!\,(l-n-1)!!}{(l+n)!!\,(l-n)!!},
  \label{eq:formfactor}
\end{equation}
$$\qquad
  F_l^{(n)}(\Delta)=\exp\Bigl[-\tfrac12\bigl(l(l+1)-n^2\bigr)\Delta^2\Bigr],$$
the form factor following from the equatorial curvature $\partial_u^2P_l^n(0)/P_l^n(0)=-[l(l+1)-n^2]$ (accurate to $2\%$ for $l\Delta\lesssim2$). The direct-term weights are $w_{nl}\propto|\Ylm ln(\pi/2,0)|^2F_l^{(n)}(\Delta)$, and
\begin{equation}
  \gamma_n^{\eff}(\Delta)=
  \frac{\sum_l|\Ylm ln(\pi/2,0)|^2F_l^{(n)}(\Delta)\gamma_l}
       {\sum_l|\Ylm ln(\pi/2,0)|^2F_l^{(n)}(\Delta)}.
  \label{eq:effspectrum}
\end{equation}
For $\Delta\gtrsim1$, $w_{nl}\to\delta_{ln}$ and the anchors of Table~\ref{tab:anchors} are recovered ($2.5000$, $4.5000$ at $\Delta=2$). For $\Delta\to0$ the sums are dominated by $l\sim1/\Delta$, where $|\Ylm ln|^2$ is $n$ independent, so $\gamma_n^\eff\to\Delta^{-2}+\mathcal O(1)$ and
\begin{equation}
  \gamma_n^\eff/\gamma_2^\eff\simeq1+(n^2-4)\Delta^2\qquad(\Delta\to0),
  \label{eq:washout}
\end{equation}
which reproduces the numerics at $\Delta=0.1$ ($1.0494$ vs.\ $1.05$; $1.1186$ vs.\ $1.12$); Table~\ref{tab:washout} gives the interpolation. Since the response is built at $t\lesssim R$, where $\Delta\sim\tau_0/R$, large systems suppress the mode-resolved signal both hydrodynamically and through the squeeze, while small systems with $\tau_0/R=\mathcal O(0.3$--$1)$ retain $\mathcal O(1)$ sensitivity. Two caveats: a bounded spectrum ($\gamma_l\to\gamma_\infty$ above $l_\theta\sim1/\theta_{\rm typ}$; cf.\ the bottom-up hierarchy \cite{Baier:2000sb}) also drives Eq.~\eqref{eq:effspectrum} toward $n$ independence, so the intermediate-$\Delta$ double ratio is sensitive to $l_\theta$; and the equilibrium-subtraction weights decay geometrically in $l$ and shift the observable back toward the anchors. Chapman--Enskog on Eq.~\eqref{eq:boltzmann3d} gives a single shear viscosity from the five $m$-degenerate $l=2$ modes, $\eta\propto(e+P)/\gamma_2$, so the tower is invisible in the hydrodynamic limit.

\begin{table}[t]
\centering
\caption{Double ratios of effective rates, Eq.~\eqref{eq:effspectrum}, for the 3D spectrum $\gamma_l\propto l(l+1)-2$ versus squeeze width $\Delta$.}
\label{tab:washout}
\begin{tabular}{lccccc}
\toprule
$\Delta$ & $1.0$ & $0.5$ & $0.3$ & $0.2$ & $0.1$ \\
\midrule
$\gamma_3^\eff/\gamma_2^\eff$ & $2.49$ & $1.93$ & $1.41$ & $1.19$ & $1.05$ \\
$\gamma_4^\eff/\gamma_2^\eff$ & $4.49$ & $3.23$ & $1.97$ & $1.46$ & $1.12$ \\
\bottomrule
\end{tabular}
\end{table}

\section{The leading-logarithmic Fokker--Planck kernel}
\label{sec:fp}

\subsection{Transport model}
\label{sec:fp-kernel}

The solver evolves $F(\xperp,\phi,p,\tau)\equiv p^2f$ under the leading-logarithmic kernel \meq{6}, whose circular-harmonic blocks are $C_l=-\Lambda_l$ with [\meq{7}]
\begin{equation}
  \Lambda_l=-\kappa\Bigl[\partial_p^2+\Bigl(\frac1p+\frac1T\Bigr)\partial_p+\frac1{pT}-\frac{l^2}{p^2}\Bigr],
  \qquad\gamma_l(p)=\frac{\kappa l^2}{p^2}.
  \label{eq:lambdal}
\end{equation}
The collision term relaxes $F_l-F_l^{\eq}$ under $\Lambda_l$ for $|l|\ge2$, multiplied by $r(\xperp,\tau)=g(e/e_{\rm ref})^\alpha(\tau_0/\tau)$; the $l=0,\pm1$ channels are frozen, so conservation is exact. [The projection $C_{\rm proj}=-(1-\mathcal P_{\rm cons})\Lambda(1-\mathcal P_{\rm cons})$, which retains the non-conserved radial dynamics of these channels, changes no result.] The Landau-matched equilibrium is $F_{\eq}=[e_{\rm LRF}/(4\pi T_0^3)]p^2e^{-p\sigma(\phi)/T_0}$, $\sigma=u^\tau-u_x\cos\phi-u_y\sin\phi$. In the benchmark the bath scale is frozen to $T_0$ (Sec.~\ref{sec:fp-localT} relaxes this), and $\kappa=T_0^2/2$ so that the thermal one-hit $l=2$ rate equals $g$, matching the $\hat g_2=1$ convention of the scalar spectra.

\subsection{Spectrum}
\label{sec:fp-spectrum}

$\Lambda_l$ is self-adjoint with respect to $w(p)=p\,e^{p/T}$. The Liouville substitution $\psi=w^{-1/2}\chi$ \cite{Vogel:1989zz}, used for relativistic Fokker--Planck kinetic theory by Gavassino \cite{Gavassino:2026zsz,Gavassino:2026tvy}, maps each block onto the radial Coulomb Hamiltonian \meq{9} with angular momentum $l-\tfrac12$, charge $1/2T$, and offset $\kappa/4T^2$. Ref.~\cite{Gavassino:2026zsz} treats the conserved-density sector; here the mapping is applied to the anisotropy blocks $l\ge2$. The spectrum is [\meq{10}]
\begin{equation}
  \gamma_{l,k}=\frac{\kappa}{4T^2}\Bigl[1-\frac1{(2l+2k+1)^2}\Bigr],\quad k=0,1,\dots,
  \label{eq:rydberg}
\end{equation}
\begin{equation}
  \gamma_{\rm cont}\ge\frac{\kappa}{4T^2},\qquad
  \psi_{l,k}\propto p^{\,l}e^{-p/2T}e^{-p/4T\nu}L_k^{(2l)}\bigl(\tfrac{p}{2T\nu}\bigr),
  \label{eq:rydbergpsi}
\end{equation}
$$\nu=l+k+\tfrac12 .$$
A fine-grid diagonalization of Eq.~\eqref{eq:lambdal} reproduces Eqs.~\eqref{eq:rydberg},~\eqref{eq:rydbergpsi} to $3\times10^{-4}$ for $l=2,3,4$, $k=0,1$, including the degeneracy $\gamma_{l,k}=\gamma_{l+1,k-1}$. The $l^2$ hierarchy resides in the continuum (soft momenta); the bound states, at hard momenta, lie between $0.96$ and $1$ times the edge, with $\gamma_{3,0}/\gamma_{2,0}=50/49\simeq1.0204$ [\meq{11}].

\subsection{Effective rate, viscosity, and breaking of factorization}
\label{sec:fp-onehit}

The one-hit expansion carries over with $\hat g_n\to\Lambda_n$ acting on the $p$ dependence of the source. For a channel-$n$ profile $h(p)$ in $f$ space [\meq{12}],
\begin{equation}
  \gamma_n^\eff[h]\equiv\frac{\langle p^2|\Lambda_n|h\rangle}{\langle p^2|h\rangle}=A[h]+n^2B[h],\qquad
  A=\kappa\frac{I_1/T-I_0}{I_2},
  \label{eq:AB}
\end{equation}
$$\quad B=\kappa\frac{I_0}{I_2},\quad I_m=\int_0^\infty\dd p\,p^mh,$$
so at leading opacity the kernel is equivalent to a mixed spectrum \eqref{eq:mixed} with $a=A[h]$, $b=B[h]$. For $h=e^{-p/T}$, $A=0$ and $\gamma_n^\eff=\kappa n^2/2T^2$, recovering $9/4$ and $4$; in the eigenbasis, $\gamma_n^\eff=\sum_kw_{nk}\gamma_{n,k}+(\text{continuum})$ with overlap weights, and the thermal average $2\kappa/T^2$ is sixteen times the edge, which is why the dilute hierarchy is quadratic although the bound spectrum is nearly degenerate.

In the $l=\pm2$ channel the Chapman--Enskog source is $\propto(p/T)e^{-p/T}$, and direct substitution gives $\Lambda_2[p^2e^{-p/T}]=(2\kappa/T)\,p\,e^{-p/T}$, so the exact solution is $\delta f_2\propto(T/2\kappa)p^2e^{-p/T}$. For a flat kernel at the thermal one-hit rate $2\kappa/T^2$, $\delta f_2^{\rm flat}\propto(T^2/2\kappa)p\,e^{-p/T}$, and with the energy weight $\int\dd p\,p^2\delta f_2$ [\meq{14}]
\begin{equation}
  \frac{\eta_{\rm FP}}{\eta_{\rm flat}}=\frac{T\int\dd p\,p^4e^{-p/T}}{T^2\int\dd p\,p^3e^{-p/T}}=4 .
  \label{eq:eta4}
\end{equation}
In units of the thermal one-hit rate, dilute response, viscosity, and late-time decay measure $1$, $1/4$, and $\gamma_{2,0}=0.12$, respectively.

Factorization breaks because the direct term keeps the thermal shape while the Landau subtraction has the harder, $n$-dependent shapes $h_n^{\eq}\propto p^2e^{-pu^\tau/T_0}I_n(p|u_\perp|/T_0)$ [\meq{13}], so $A$, $B$ are not channel universal. At one hit the direct-channel effective eigenvalues extracted from the solver are $0.9900$ and $2.2216$ ($n=2,3$), equal to the discrete-operator values of Eq.~\eqref{eq:AB} to seven digits; the subtraction-channel values are $0.279$ and $0.346$ (ratio $1.24$, close to the inverse-type average $1/4$). The total dilute anchors are $D_{32}^{\rm direct}=9/4$ exactly, $D_{32}^{\rm total}\simeq1.1$--$1.2$ at $\alpha=1$ (subtraction dominated by Lemma S~\ref{prop:evenodd}; $1.19$ at $\tau_{\max}=6R$ vs.\ $1.07$ at $4R$), $D_{32}^{\rm total}=1.54$ at $\alpha=1/3$, and $D_{42}^{\rm total}=2.77$, whereas any scalar spectrum returns $D_{32}=2.250000$ independently of $\alpha$, $\tau_{\max}$, and grid.

\subsection{Opacity dependence and momentum-space tomography}
\label{sec:fp-crossover}

With the bath frozen, the mean momentum of the $V_2$-carrying profile rises from $3T_0$ to $5.8T_0$ ($g=0.5$) and $5.3T_0$ ($g=2$) by $\tau=6R$ [\mfig{2(b)}], and $\kappa_2^{\rm FP}/\kappa_2^{\rm flat}=0.25\to0.11$ across $g=0.5\to8$, the response-side counterpart of Eq.~\eqref{eq:eta4}. The double ratio is given in Table~\ref{tab:d32fp} ($p_{\min}=0$, so $g\equiv\chi_2$ of Eq.~\eqref{eq:chi2}).
 For the Fokker--Planck kernel the triangular response changes sign at $g_3^*\in(2,3)$, beyond which the double ratio is ill-conditioned.

\begin{table}[ht]
\centering
\caption{Double ratio $D_{32}(g)$ for angular diffusion and for the Fokker--Planck kernel (frozen bath, $\alpha=1$).}
\label{tab:d32fp}
\begin{tabular}{lccccc}
\toprule
 & $g=0.5$ & $g=1$ & $g=2$ & $g=4$ & $g=8$ \\
\midrule
$D_{32}$, & & & & & \\
separable $\gamma_l\propto l^2$      & $2.230$ & $2.208$ & $2.163$ & $2.075$ & $1.922$ \\
\addlinespace 
$D_{32}$, & & & & & \\
F-P kernel, frozen bath   & $1.600$ & $1.846$ & $0.73$  & \multicolumn{2}{c}{$\kappa_3<0$} \\
\bottomrule
\end{tabular}
\end{table}

The triangular response changes sign at $g_3^*\in(2,3)$ ($\kappa_3=+6.6\times10^{-4},\,-4.1\times10^{-4},\,-1.5\times10^{-3}$ at $g=2,3,4$), because the positive subtraction-generated and the negative direct contributions damp at different effective rates; every scalar spectrum keeps $\kappa_3>0$. The crossing is a property of the frozen-bath model (Sec.~\ref{sec:fp-localT}).

Bin-resolved rates for $p\in[0,2),[2,4),[4,8)\,T_0$ follow from the eigenfactorization for a uniform system, $\gamma_n^{\eff,\rm bin}(t)=-\dd\ln V_n^{\rm bin}/\dd t$, $V_n^{\rm bin}=\mathbb 1_{\rm bin}^{\top}e^{-\tilde\Lambda_nt}h$ [\mfig{1(a)}]:
\begin{center}
\small
\begin{tabular}{c|ccc}
\toprule
$t=g\!\int\!r\,\dd\tau$ & soft $[0,2)$ & mid $[2,4)$ & hard $[4,8)$ \\
\midrule
$0$      & $2.250$ & $2.250$ & $2.249$ \\
$0.5$    & $1.601$ & $2.004$ & $2.226$ \\
$2$      & $1.424$ & $1.548$ & $1.877$ \\
$8$      & $1.299$ & $1.318$ & $1.372$ \\
$\infty$ & \multicolumn{3}{c}{$\gamma_{3,0}/\gamma_{2,0}=1.0204$} \\
\bottomrule
\end{tabular}
\end{center}
At one hit $\Lambda_ne^{-p/T}=(\kappa n^2/p^2)e^{-p/T}$, so every bin starts at $9/4$ (verified to $2\times10^{-4}$). A bin displays the modes that survive in it: at soft momenta the continuum is removed first, leaving the $p^{\,l}$ tails of the near-degenerate bound states, so the soft bin approaches $50/49$ first while the hard bin retains $9/4$ longest. The bin-resolved response [\mfig{1(b)}] has $\kappa_2^{\rm soft}=-7\times10^{-3}$, $-4.0\times10^{-2}$ and $\kappa_2^{\rm hard}=+7.0\times10^{-3}$, $+2.8\times10^{-2}$ at $g=2,8$; a scalar kernel cannot produce a momentum-differential sign structure.

\subsection{Conformal rate exponent}
\label{sec:fp-alpha13}
\begin{figure*}[t]
\centering
\includegraphics[width=\linewidth]{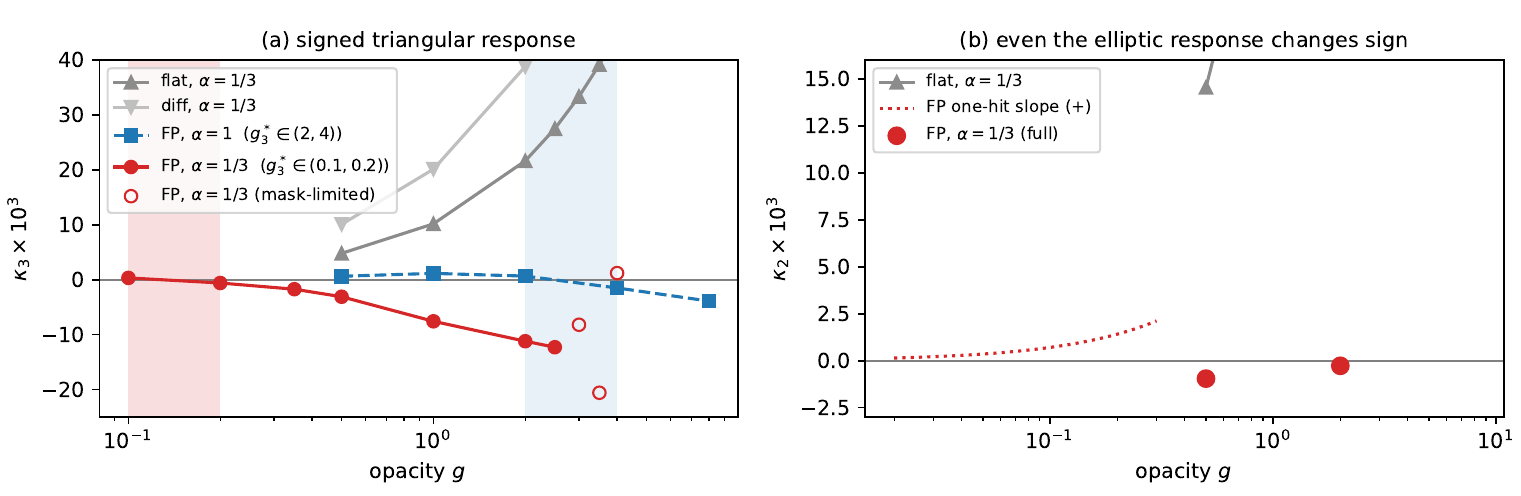}
\caption{Signed response coefficients at $r\propto e^{1/3}$, frozen bath. (a) $\kappa_3(g)$: Fokker--Planck kernel (circles; open symbols mask limited) crosses zero in the red band, earlier than at $\alpha=1$ (squares, blue band); scalar spectra (grey) remain positive. (b) $\kappa_2(g)$: the Fokker--Planck response passes from the positive one-hit slope (dotted) to negative values.}
\label{fig:alpha13}
\end{figure*}

At $\alpha=1/3$ the odd direct channel is restored and negative ($-6.6\times10^{-4}$ against a subtraction of $+3.9\times10^{-3}$ at one hit). The frozen-bath scan at thirteen opacities $g=0.1$--$8$ (Fig.~\ref{fig:alpha13}) shows the triangular crossing at $g_3^*\in(0.1,0.2)$ with $\kappa_3$ reaching $-1.2\times10^{-2}$ by $g\simeq2$, while $\kappa_3^{\rm flat}=+2.2\times10^{-2}$ ($g=2$), $+8.8\times10^{-2}$ ($g=8$). The direct $n=2$ term is also negative ($-5.4\times10^{-4}$ at one hit): $\kappa_2$ crosses zero below $g=0.5$, equals $-0.8\times10^{-3}$ there, and recovers by $g=2$, against $+1.5\times10^{-2}$ for scalar kernels. Because $e^{1/3}$ decays slowly, the halo carries opacity: the one-hit limit is reached at $g=0.05$ to $6\%$, box doubling ($L=12$, $N_x=108$) moves $\kappa_3(g{=}0.5)$ by $10\%$, cutoff refinement $10^{-7}\to10^{-9}e_{\rm ref}$ moves $\kappa_3(g{=}2)$ by $2\%$; the scan is converged for $g\lesssim2.5$, and deeper points are marked resolution limited.

\subsection{Local temperature}
\label{sec:fp-localT}

The benchmark of Secs.~\ref{sec:fp-crossover}- \ref{sec:fp-alpha13} freezes the temperature that enters the operator \eqref{eq:lambdal} and the Landau
equilibrium to the reference value $T_0$. In an expanding fireball this
temperature falls with the local energy density, and one must ask which of
the benchmark results survive when it does. The question can be separated
into two parts because the Fokker--Planck operator is self-similar under a
rescaling of the momentum axis: with $(Sf)(p)=f(pT/T_0)$,
\begin{equation}
  \Lambda_l^{(T)}=\frac{T}{T_0}\,S^{-1}\Lambda_l^{(T_0)}S .
  \label{eq:selfsimilar}
\end{equation}
A change of temperature therefore does two things. It multiplies the whole
operator by $T/T_0$, and it stretches the momentum profile on which the
operator acts. The first effect changes only the overall rate: the entire
tower \eqref{eq:rydberg} scales as $\gamma_{l,k}\propto T$, the eigenfunctions
depend on $p/T$ alone, and the viscosity ratio \eqref{eq:eta4} holds at any
uniform $T$. For a conformal transport coefficient, $\kappa\propto T^3$, the
factor $T/T_0=(e/e_{\rm ref})^{1/3}$ is exactly a shift $\alpha\to\alpha+\tfrac13$
of the rate exponent, which the scan of Sec.~\ref{sec:fp-alpha13} already
covers. The second effect, the rescaling of the shape, is the genuinely new
one, and it is isolated by holding the rate normalization $r(\xperp,\tau)$ of
the frozen model fixed while evaluating both the operator and the Landau
equilibrium at the conformal local temperature
\begin{equation}
  T(\xperp,\tau)=T_0\bigl(e_{\rm LRF}/e_{\rm ref}\bigr)^{1/3}.
  \label{eq:localT}
\end{equation}
Numerically this is implemented through Eq.~\eqref{eq:selfsimilar} as a
per-point resampling of the collisional change, which is exact at zero rate,
reduces to the identity at uniform $T$ to $2\times10^{-14}$, and conserves
energy and momentum exactly. The radial grid is $N_p=32$, and a floor
$T\ge0.6T_0$ is applied in the energy-suppressed halo (lowering it to
$0.5T_0$ moves $\kappa_3$ by $8\%$). A scan at $g=1$--$4$ for $n=2,3$ then
gives the following [\mfig{2}].

The momentum hardening survives and becomes stronger. With the bath cooler
over most of the fireball, the flow-carrying deviation is harder relative to
the local scale: the mean momentum of the $V_2$-carrying profile at $g=2$
reaches $8.5T_0$ by $\tau=6R$, against $5.3T_0$ for the frozen bath.

The separation from scalar kernels survives. The double ratio settles on a
plateau, $D_{32}=1.475,\,1.505,\,1.543$ at $g=1,2,4$, well separated from the
flat baseline ($D_{32}=1$), from angular diffusion ($2.21$--$2.07$ over the
same range), and from the frozen-bath trajectory. The value is the
mixed-spectrum ratio $(a+9b)/(a+4b)=3/2$ at $b/a=1/6$, Eq.~\eqref{eq:mixed}:
the response lands at a point that no single-rate or pure-diffusion spectrum
produces, as the effective-rate decomposition \eqref{eq:AB} predicts for a
source with both a soft and a hard component.

The zero crossing does not survive. With the local temperature $\kappa_3$ is
positive and increases monotonically through $g=4$
($+6.9,\,+14.1,\,+21.1,\,+27.7\times10^{-3}$), at both rate exponents; the
fully conformal model (local $T$ together with $\alpha=1/3$) gives
$\kappa_3=+5.2,\,+13.7,\,+49.0\times10^{-3}$ at $g=0.2,\,0.5,\,2$, where the
frozen bath was already negative. The cooler bath damps the
subtraction-generated flow less, so the cancellation that produced the sign
change in the frozen model never completes.

The sign changes of Secs.~\ref{sec:fp-crossover} and \ref{sec:fp-alpha13}
are therefore properties of the frozen-bath rate model. The results that do
not depend on that choice are the eigenvalue tower and its $T$ scaling, the
viscosity ratio, the channel-resolved effective eigenvalues, the momentum
hardening, and the mixed-spectrum plateau of $D_{32}$.

\subsection{Validation and infrared robustness}
\label{sec:fp-numerics}

The momentum-resolved solver uses the same exact spectral advection and the exact operator exponential per harmonic, $e^{-r\Delta\tau\Lambda_l}=R_le^{-r\Delta\tau\Gamma_l}L_l$, from a conservative Sturm--Liouville discretization of Eq.~\eqref{eq:lambdal}. Tests: a scalar spectrum through the same pipeline reproduces the $p$-integrated solver ($|\Delta V_n|<10^{-4}$) and $D_{32}=2.250000$; seeded eigenmodes decay at their analytic rates to $10^{-14}$; energy and momentum conserved to $10^{-14}$; one-hit agreement to $<1\%$ at $g=0.02$ in both $n=2$ and $n=3$; $V_n(-\beta)=-V_n(\beta)$ to $10^{-13}$. Production grids $N_x=80$, $N_\phi=32$, $N_p=20$, $p_{\max}=12T_0$; away from the crossing, $N_p\to26$, $p_{\max}\to14$, $N_x\to96$ move $\kappa_2$ by $\le9\%$, $\le0.4\%$; the $g=2$ double ratio carries $D_{32}=0.73^{+0.00}_{-0.09}$ without affecting $g_3^*\in(2,4)$.

\begin{figure*}[t]
\centering
\includegraphics[width=\linewidth]{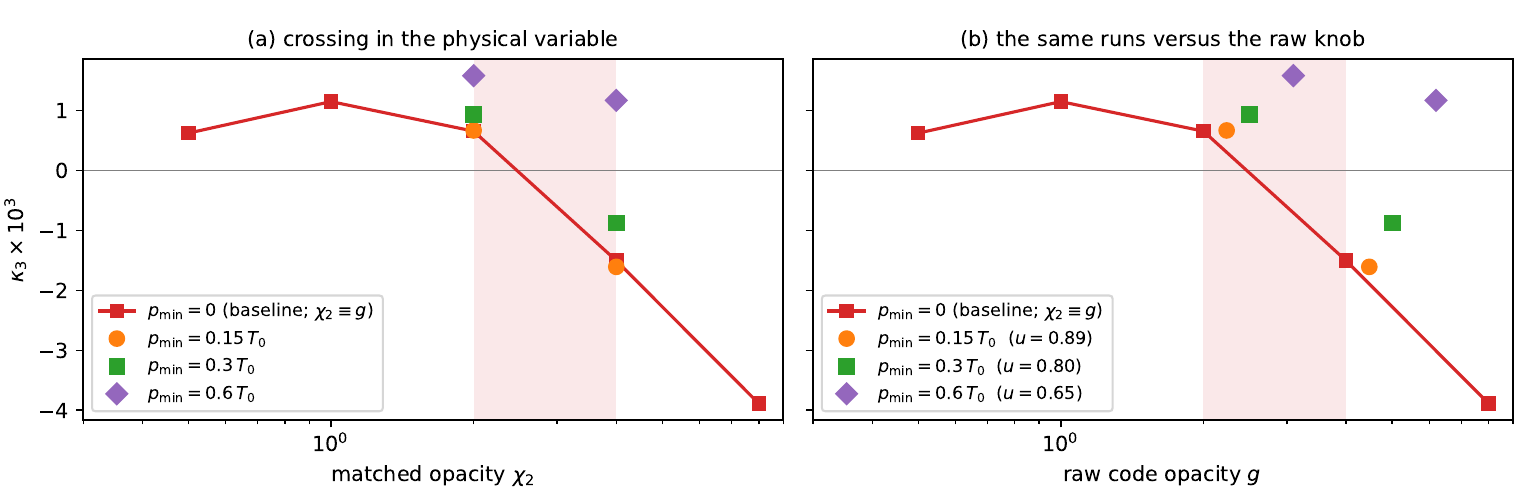}
\caption{Triangular zero crossing vs the matched opacity $\chi_2$, Eq.~\eqref{eq:chi2}. (a) $\kappa_3$ versus $\chi_2$: cutoff configurations rerun at $\chi_2=2,4$ collapse onto the $p_{\min}=0$ baseline and cross zero in a common window (shaded); the residual shift at $p_{\min}=0.6T_0$ is the channel factor $\rho(a)$. (b) The same runs versus the raw parameter $g$.}
\label{fig:chi2}
\end{figure*}

The $p\to0$ singularity of $\gamma_l(p)$ is integrable under $p^2\dd p$. With a cutoff $p_{\min}=a$, $x=a/T$,
\begin{equation}
  \gamma_n^\eff(a)=\kappa\frac{n^2T+a}{a^2+2aT+2T^2},\qquad
  \frac{D_{32}^{\rm direct}(a)}{9/4}=\frac{1+x/9}{1+x/4},
  \label{eq:ircut}
\end{equation}
a drift of $-3.9\%$ at $x=0.3$. The bound states shift only at $\mathcal O((a/2T\nu)^{2l})$, and the viscosity ratio at fixed physical normalization is $4+\mathcal O(a^4)$; the apparent drift $X(a)=4[1-\tfrac34x+\dots]$ is the renormalization of the opacity unit $\gamma_2^\eff(a)$. The regulator-invariant variable is the matched opacity
\begin{equation}
  \chi_2\equiv\int_{\tau_0}^{\tau_f}\dd\tau\,\overline\gamma_2^{\,\rm ref}(\tau)
  \propto g\,\frac{\gamma_2^\eff(p_{\min})}{\gamma_2^\eff(0)},
  \label{eq:chi2}
\end{equation}
with $\chi_2=g$ at $p_{\min}=0$. Table~\ref{tab:ir} and Fig.~\ref{fig:chi2}: at $\chi_2=2$, $\kappa_3=+6.69\times10^{-4}$ ($p_{\min}=0.15T_0$) vs.\ $+6.57\times10^{-4}$ ($p_{\min}=0$); $\chi_2^*\in(2,4)$ for all variations of $N_p$, $p_{\max}$, inner boundary condition, and $p_{\min}\le0.3T_0$; for $a=0.6T_0$ (removing $26\%$ of $I_0$) the residual $\chi_2^*\in(4,5.2)$ is the predicted channel-differential factor $\rho(a)=[\gamma_3^\eff(a)/\gamma_3^\eff(0)]/[\gamma_2^\eff(a)/\gamma_2^\eff(0)]=4(9+x)/[9(4+x)]=0.93$. The ground-state eigenvalue is $\gamma_{2,0}=0.1200134$ in all rows, and the numerical $X$ tracks Eq.~\eqref{eq:ircut} to $3\times10^{-3}$.

\begin{table*}[t]
\caption{Infrared and resolution tests (frozen bath, $\alpha=1$). $X$: measured viscosity ratio in units of $\gamma_2^\eff(a)$; $\chi_2^*$: triangular zero crossing in the matched variable. $\dagger$: rerun at $g=\chi_2\gamma_2^\eff(0)/\gamma_2^\eff(a)$ so that $\kappa_3$ is probed at $\chi_2=2,4$; the Dirichlet row is the rescaled raw bracket.}
\label{tab:ir}
\begin{center}
\small
\begin{tabular}{cccc|cc|c|cc}
\toprule
$p_{\min}/T_0$ & $N_p$ & $p_{\max}/T_0$ & inner BC & $X$ (num.) & $X$ [Eq.~\eqref{eq:ircut}] & $D_{32}(\chi_2{=}2)$ & $g_3^*$ & $\chi_2^*$ \\
\midrule
$0$    & $20$ & $12$ & exact     & $4.000$ & $4.000$ & $+0.73$ & $(2,4)$ & $(2,4)$ \\
$0.15$ & $20$ & $12$ & no-flux   & $3.574$ & $3.574$ & $+0.58$ & $(2,4)$ & $(2,4)^{\dagger}$ \\
$0.30$ & $20$ & $12$ & no-flux   & $3.199$ & $3.198$ & $+0.66$ & $(2,4)$ & $(2,4)^{\dagger}$ \\
$0.30$ & $20$ & $12$ & Dirichlet & $3.196$ & $3.198$ & --      & $(2,4)$ & $(1.6,3.2)$ \\
$0.60$ & $19$ & $12$ & no-flux   & $2.601$ & $2.592$ & $+0.83$ & $(4,8)$ & $(4,5.2)^{\dagger}$ \\
$0$    & $14$ & $12$ & exact     & $4.000$ & $4.000$ & $+0.30$ & $(2,4)$ & $(2,4)$ \\
$0$    & $20$ & $10$ & exact     & $4.000$ & $4.000$ & $+0.89$ & $(2,4)$ & $(2,4)$ \\
$0$    & $26$ & $14$ & exact     & $4.000$ & $4.000$ & $+0.65$ & $(2,4)$ & $(2,4)$ \\
\bottomrule
\end{tabular}
\end{center}
\end{table*}

The solver \textsc{SolKin}, with both pipelines, the validation tests, and the scripts producing all figures and tables, is archived on Zenodo \cite{Soloveva:2026SolKin}.

\bibliographystyle{apsrev4-2}
\bibliography{rtamodes}